\documentclass[11pt,a4paper]{article}
\pdfoutput=1
\usepackage{jcappub}
\usepackage{hyperref}
\hypersetup{colorlinks=true,linkcolor=blue,citecolor=blue,urlcolor=blue}

\usepackage{siunitx}
\DeclareSIUnit{\parsec}{pc}

\usepackage{amsmath,amssymb,amsthm,cleveref,aas_macros}

\usepackage{bm}
\newcommand{\bb}[1]{\bm{\mathrm{#1}}}

\usepackage[sort&compress]{natbib}
\usepackage[dvipsnames]{xcolor} % must load before tikz
\usepackage{enumerate}

\newcommand{\du}{\mathrm{d}}
\newcommand{\rhonfw}{\rho_{\mathrm{NFW}}}
\newcommand{\rhoexp}{\rho_{\mathrm{exp}}}
\newcommand{\rtidal}{r_b}
\renewcommand\u[1]{{\mathrm{#1}}}
\newcommand{\us}[1]{~{\mathrm{#1}}}
\newcommand{\ee}[1]{\times10^{{#1}}}
\newcommand{\citeil}[1]{Ref.~\cite{#1}}

\newcommand{\revised}[1]{#1}                   % clean

\title{Connecting direct and indirect detection with a dark spike in the cosmic-ray electron spectrum}
\author[a,b,c]{Adam Coogan,}
\emailAdd{a.m.coogan@uva.nl}
\author[b,c]{Benjamin V. Lehmann,}
\emailAdd{blehmann@ucsc.edu}
\author[b,c]{and Stefano Profumo}
\emailAdd{profumo@ucsc.edu}

\affiliation[a]{Gravitation Astroparticle Physics Amsterdam (GRAPPA),\\
Institute for Theoretical Physics Amsterdam
and Delta Institute for Theoretical Physics,\\
University of Amsterdam, Science Park 904, 1098 XH Amsterdam, The Netherlands}
\affiliation[b]{Department of Physics, University of California Santa Cruz,\\
1156 High St., Santa Cruz, CA 95064, USA}
\affiliation[c]{Santa Cruz Institute for Particle Physics,\\
1156 High St., Santa Cruz, CA 95064, USA}

\abstract{
Multiple space-borne cosmic ray detectors have detected line-like features in the electron and positron spectra. Most recently, the DAMPE collaboration reported the existence of such a feature at \SI{1.4}{\tera\electronvolt}, sparking interest in a potential dark matter origin. Such quasi-monochromatic features, virtually free of any astrophysical background, could be explained by the annihilation of dark matter particles in a nearby dark matter clump. Here, we explore the consistency of producing such spectral features with dark matter annihilation from the standpoint of dark matter substructure statistics, constraints from anisotropy, and constraints from gamma-ray emission. We demonstrate that if indeed a high-energy, line-like feature in the electron-positron spectrum originates from dark matter annihilation in a nearby clump, a significant or even dominant fraction of the dark matter in the Solar System likely stems from the clump, with dramatic consequences for direct dark matter searches.
}

\begin{document}
\maketitle

\section{Introduction}
\label{sec:introduction}
The DArk Matter Particle Explorer (DAMPE), successfully launched at the end of 2015, is a high-energy particle detector with a large acceptance of ${\cal O}(0.3\ {\rm m}^2~{\rm sr})$ and a deep calorimeter ($\sim$32 radiation lengths) designed to measure cosmic-ray electrons and positrons (CRE) and gamma rays up to energies greater than \SI{10}{\tera\electronvolt} \cite{Ambrosi:2017wek}. The detector has impressive energy resolution (better than 1.2\% at energies greater than \SI{100}{\giga\electronvolt}) and a robust electron/proton discriminating power \cite{Ambrosi:2017wek}. Recently, the DAMPE collaboration reported a new measurement of the CRE flux in the energy range between \SI{25}{\giga\electronvolt} and \SI{4.6}{\tera\electronvolt} taken during more than 500 days of data-taking operations. The data bridges the gap between data taken with smaller space-borne detectors, such as AMS-02 \cite{Aguilar:2013qda} and Fermi LAT \cite{Ackermann:2010ij}, and with ground-based Cherenkov data, e.g. HESS. \cite{Aharonian:2008aa}. 

One of the key results of the DAMPE CRE flux measurement is a firm, highly-significant confirmation of a spectral break at energies around a TeV, consistent with previous measurements at lower \cite{Aguilar:2013qda, Ackermann:2010ij} and higher \cite{Aharonian:2008aa} energy. Due to efficient CRE energy losses at high energies, the relevant spatial region where sources of TeV CRE can exist is limited to within a sphere of around \SI{1}{\kilo\parsec} the Solar System. The TeV break is then likely due to the transition to a high-energy regime where the CRE is dominated by a few local sources, with injection spectra approaching a high-energy exponential cut-off \cite[e.g.][]{Profumo:2008ms, DiMauro:2014iia}.

A second intriguing feature of the CRE spectrum measured by DAMPE is a rather prominent line-like feature at \SI{1.4}{\tera\electronvolt}, limited to an excess of counts in a single energy bin \cite{Ambrosi:2017wek}. The statistical significance for the line-like excess is low with current data: the independent analysis of \citeil{Fowlie:2017fya} finds a local significance of 3.6$\sigma$ and a global significance, including the look-elsewhere effect, of 2.3$\sigma$. Additionally, the CALET experiment onboard the International Space Station recently reported a measurement of the CRE flux in the energy range between \SI{10.6}{\giga\electronvolt} and \SI{4.75}{\tera\electronvolt} \cite{Adriani:2018ktz}. While CALET data confirm the spectral break observed by DAMPE, the measured CRE flux is not compatible with the \SI{1.4}{\tera\electronvolt} DAMPE bin at a level of 4$\sigma$ significance, including systematic errors from both CALET and DAMPE and possible binning effects \cite{Adriani:2018ktz}. Nevertheless, the ``DAMPE excess'' has generated significant attention, and \revised{possible origins of} the ``line'' have been identified both in the realm of new physics beyond the standard model, and in astrophysical sources such as pulsars. 

In contrast to the positron excess, which covers more than a decade in energy, a line-like feature seems rather unlikely to originate from pulsars: in particular, the injection spectrum must be extraordinarily ``cold'' \cite{Bogovalov:2000pu, 1984ApJ...283..694K}, meaning extremely narrow in energy, and unaffected by the supernova remnant shock surrounding the neutron star. In addition, the candidate source must be quite close, \revised{possibly sourced by} the nearby pulsars Geminga and Vela \cite[see e.g.][]{Yuan:2017ysv}. Whether pulsar injection spectra can indeed possess an energy spectrum with a feature exhibiting $\Delta E/E\sim 5\%$ is theoretically and observationally unproven. Such a feature has \revised{not been observed} in direct measurements of the CRE injected by local pulsars, such as Geminga itself and Monogem \cite{Abeysekara:2017old, Profumo:2018fmz}.

Thus, significant attention has been given to the possibility of a dark matter \revised{origin of} the DAMPE excess. The fit of a given dark matter model is dependent upon the spectral shape of the excess, as we discuss and quantify in detail below. Very high-energy CRE lose energy very efficiently via both synchrotron and inverse Compton emission, with $\du E/\du t\sim E^2$. As a result, the injection spectrum, i.e. the spectrum of the CRE resulting from dark matter annihilation or decay, should be very close to monochromatic. One possibility is that the dark matter is ``electrophilic,'' so that an annihilation or decay event only (or largely) produces monochromatic electrons and positrons. Here, we will assume that $\chi\chi\to e^+e^-$ in 100\% of the annihilation events. Secondly, the source must be very close, or ``old'' CRE will diffuse for times long enough for them to populate lower-energy bins to a degree conflicting with data. We note that dark matter decay is strongly disfavored given constraints from secondary emission from inverse Compton \cite[see e.g.][]{Dugger:2010ys} as well as from diffuse gamma-ray searches, which we discuss in detail below. We will concentrate on the possibility of annihilation for the remainder of this work.

A nearby, high-density source of dark matter annihilation would require very large over-densities, likely too large to be statistically viable. Alternately, large densities could originate from intermediate-mass black holes surrounded by a dark matter density spike formed through adiabatic accretion \cite{Zhao:2005zr}, or from ultracompact minihalos \cite{Bringmann:2011ut}. While rather speculative, both possibilities might be advocated to attain a viable source structure for the \revised{origin of} a line-like feature in the CRE \cite[see e.g.][]{Yang:2017cjm}. 

Here, we contemplate the possibility that the dark matter structure responsible for possibly sourcing the DAMPE line-like feature at \SI{1.4}{\tera\electronvolt} is a dark matter ``clump,'' with some characteristic size and at some distance from Earth. We discuss constraints on such a scenario and evaluate the probability of finding such a structure nearby. Our setup captures as extreme cases all possibilities discussed above, including that of mini-spikes and that of ultracompact minihalos. However, we note that the probabilities we calculate apply only to a standard scenario where no enhancement is caused by adiabatic accretion on a massive body. Rather, we utilize results of $N$-body dark-matter-only simulations to compute the likelihood of finding an adequately ``luminous'' clump.

In this study we consider a variety of constraints on the properties of possible dark matter structures sourcing the DAMPE excess. We calculate both the gamma-ray flux and the CRE emission fully including the possibly {\em extremely extended structure} of the dark matter source, and quantify the difference between this complete calculation and the assumption of a point-like source. We find that such difference is crucial: for example, constraints from gamma rays are generally significantly weakened, and extended CRE sources \revised{constraints from anisotropies in CRE arrival directions are less stringent} (see e.g. the general result in \citeil{Profumo:2014yxa} for a point-like source).

Our central result is a somewhat unexpected {\em connection of indirect and direct dark matter detection}: we show here that  should a feature indeed be present in the CRE spectrum, such as the DAMPE line, rather generically one would expect a significant contribution of dark matter at Earth from the dark matter source of the CRE excess. While we find that the relative contribution to the local dark matter density can be small (of order a percent), in the most likely scenarios it is an enhancement of a factor 10 or more. This has profound implications for direct detection rates: first, the rate scales linearly with the local dark matter density. Thus, if a CRE line originates from dark matter, it would likely mean much larger direct detection rates are expected. Additionally, since the velocity distribution of a small, nearby clump is likely very narrow, if the clump is close to virialized, then the velocity distribution of dark matter particles will inevitably contain a cold stream, with important and conspicuous consequences for the {\em differential} event rate at direct dark matter searches. In light of our results, and even disregarding this latter effect, \revised{null results at direct detection experiments may imply much more stringent bounds than typically drawn. Moreover, in this scenario, the experimental sensitivity to dark matter is enhanced relative to neutrino scattering, so the neutrino background becomes important only at smaller values of the dark matter-baryon cross section. In fact,} direct dark matter detection might already be \revised{probing cross sections} beyond the so-called neutrino floor \cite{Billard:2013qya}, \revised{as projected without an enhancement to the local dark matter density.}

The structure of this study is as follows: in~\cref{sec:dmclumps}, we discuss the dark matter density profiles of the clumps under consideration. In \cref{sec:cre}, we describe the production of CRE from a nearby dark matter clump, and what a signal like that possibly found by DAMPE would imply, including constraints from the signal width and from CRE anisotropy. \Cref{sec:gammas} describes the calculation of the gamma-ray spectrum from the (extended) clump, and resulting constraints from point source searches and diffuse emission. \Cref{sec:clumpstats} compares the properties of the clumps needed to explain a DAMPE-like signal with the expected probability of finding such a clump, extrapolating the latter from the results of $N$-body simulations. \Cref{sec:results} presents our results in the context of the DAMPE excess for a variety of assumed clump ``shapes.'' We summarize and conclude in~\cref{sec:conclusions}.

\section{Density profiles of dark matter clumps}\label{sec:dmclumps}
Dark matter substructures (clumps) are an ideal target for indirect detection searches. The luminosity due to dark matter self-annihilation scales with the square of the dark matter density, so a relatively small clump can source a bright annihilation feature if it has a sufficiently steep density profile. Further, the abundance and spatial distribution of such clumps are predicted by $N$-body simulations, so it is possible to estimate the statistics of the nearby substructure population. In this section, we introduce the clump profiles and parametrizations that we use to compute the flux of annihilation products and describe their parameter distributions.

The structure of dark matter halos is often modeled by the Navarro-Frenk-White~\citep[NFW, ][]{Navarro:1995iw} density profile, which provides a good fit to simulated halos with a wide range of sizes. We use a three-parameter generalization of the NFW profile with the form
\begin{align}
    \label{eq:nfw}
    \rhonfw(r) &= \rho_0\left( \frac{r}{r_s} \right)^{-\gamma} \left( 1 + \frac{r}{r_s} \right)^{\gamma-3}.
\end{align}
Here $\gamma$ controls the steepness of the inner density profile, which goes as $\rho(r)\sim r^{-\gamma}$ for $r\ll r_s$. The standard NFW profile is recovered for $\gamma=1$. The parameter $\rho_0$ characterizes the overdensity of the halo relative to the background.

The third parameter, $r_s$, is known as the scale radius, and it controls the spatial extent of the halo. While $r_s$ controls the size of the halo, it does not bound the halo---rather, the profile extends to infinity. However, it is necessary for comparison with $N$-body simulations to define a strict boundary, since the mass contained in an NFW halo is divergent otherwise. There are several boundary definitions in common use. The virial radius, $r_{\mathrm{vir}}$, is defined such that the average density within $r_{\mathrm{vir}}$ is larger than the critical density by a factor $\Delta$. The virial mass, $M_{\mathrm{vir}}$, is then defined as the mass contained within the virial radius. In this work, we use $\Delta=200$ when working with the virial radius and mass. In the literature, with this choice of $\Delta$, $r_{\mathrm{vir}}$ and $M_{\mathrm{vir}}$ are sometimes denoted by $r_{200}$ and $M_{200}$ respectively. The ratio $c\equiv r_{\mathrm{vir}}/r_s$ is called the \emph{concentration} of the halo. At fixed $\gamma$, a halo is fully described either by $(r_s,\rho_0)$ or $(c,M_{\mathrm{vir}})$.

However, the virial radius is inappropriate for describing substructure, since particles this distant from the center of a clump are typically stripped away by tidal forces during the clump's evolution~\citep{Kazantzidis:2003hb,Hooper:2016cld}. As such, the subhalo masses $M_{\mathrm{sub}}$ measured in $N$-body simulations cannot be identified with $M_{\mathrm{vir}}$. For our purposes, this is significant because the mass function of nearby subhalos is computed in simulations---i.e., the mass function is the distribution of $M_{\mathrm{sub}}$ rather than $M_{\mathrm{vir}}$. Thus, when working with NFW halo masses, we must choose a boundary radius $r_{\mathrm{sub}}$ for the halo that corresponds to the boundaries of the simulated halos. We estimate $r_{\mathrm{sub}}$ by requiring that $\rho(r_{\mathrm{sub}})=2\rho_{\mathrm{host}}$, where $\rho_{\mathrm{host}}\sim\SI{0.3}{\giga\electronvolt/\centi\meter^3}$ is the local density of the Milky Way halo \cite{localdm}. This is similar to the procedure used to define subhalo masses in the Via Lactea simulation~\citep{Diemand:2007}. With $r_{\mathrm{sub}}$ and $M_{\mathrm{sub}}$ defined in this way, we can determine the distribution of $M_{\mathrm{sub}}$ from $N$-body simulations. The distribution of $r_s$ can be inferred from the distribution of halo concentrations at fixed mass. The latter has been studied in depth by e.g. \citeil{Sanchez-Conde:2014}.

Another approach to working with tidally-stripped halos is to abandon the NFW profile altogether. In particular, \citeil{Hooper:2016cld} consider a power-law profile with an exponential cutoff. We consider a modified parametrization of the same profile, with the form
\begin{equation}
    \label{eq:exp}
    \rhoexp(r) = \rho_0\left(\frac{r}{\rtidal}\right)^{-\gamma} \exp\left( -\frac{r}{\rtidal} \right).
\end{equation}
Here $\rtidal$, the tidal radius, is the effective boundary of the halo. We will refer to this tidally-truncated density profile as the ``exponential'' profile. The local population of subhalos was studied in detail by \citeil{Hooper:2016cld}, who found that the exponential profile provides a better fit to simulations, even when 99\% of the subhalo's initial mass has been tidally stripped. Moreover, \citeil{Hooper:2016cld} determines the distribution of $\gamma$ for simulated subhalos, and finds it to be approximately independent of the clump mass, with an average value $\left\langle\gamma\right\rangle=0.74$. They also determine the distribution of $\rtidal$ as a function of mass.

For the exponential profile, as in the NFW case, $\gamma$ controls the steepness of the inner density profile. It is important to note that the interior structure of dark matter clumps on small scales is poorly understood due to the limitations of simulations \citep{Hooper:2016cld,Kuhlen:2008}. Since the rate of DM annihilation is proportional to $\rho^2$, the innermost region makes a significant contribution to the total luminosity, so any calculation of the flux is quite sensitive to $\gamma$. On the other hand, the effect on direct detection is most sensitive to the radii, $r_s$ and $\rtidal$.

In this work, for concreteness, we consider both NFW and exponential density profiles, with several values of $\gamma$ for each. For the exponential case, we fix $\gamma$ to its 25th, 50th, and 75th percentile values: $0.52$, $0.74$, and $1.08$, respectively. For the generalized NFW profile, we consider $\gamma=1$ (standard NFW) as well as $\gamma=0.5$ in order to facilitate comparison with \citeil{Ge:2017tkd}. Note that since the flux is dominantly produced in the overdense inner region of the halo, a generalized NFW profile and an exponential profile with the same values of $\gamma$ and $\rho_0$ are very similar for the purposes of indirect detection.

Our focus on the NFW and exponential profiles is motivated by the properties of dark matter clustering in $N$-body simulations, but in principle, there are other dark matter structures that may exhibit comparable overdensities. Specifically, ultracompact minihalos (UCMH) or dark matter cusps around intermediate-mass black holes (IMBH) could also produce annihilation features. Thus, we separately consider a UCMH-like profile with a steep inner power law. We choose two representative values for the power law index: $\gamma=9/4$, as in \citeil{Ricotti:2009}, and $\gamma=3/2$, corresponding to the profile predicted to exist in the immediate vicinity of a central black hole~\citep{Lacki:2010}. While the abundance and properties of such objects are subject to speculation, we can still determine the characteristic effects of a UCMH on the cosmic ray spectrum, the gamma ray spectrum, and the local dark matter density.

There are two additional considerations in setting the form of such a profile. First, in place of a scale radius or a tidal radius, we truncate the profile beyond a truncation radius $r_\u{tr}$, which is left as a free parameter in our analysis. Second, the luminosity of a power law profile with index $9/4$ is divergent, so we also define a plateau radius $r_p$ such that the density is constant for $r<r_p$. The profile then has the form
\begin{equation}
    \label{eq:ucmhprofile}
    \rho_{\mathrm{UCMH}}(r)=\begin{cases}
        \rho_{\mathrm{max}}
            & r < r_p \\
        \rho_{\mathrm{max}}\left(r/r_p\right)^{-\gamma}
            & r_p \leq r < r_\u{tr} \\
        0
            & r \geq r_\u{tr}
    \end{cases}
\end{equation}
where $\rho_{\mathrm{max}}$ is a density scale set by the physics that controls the clump's formation. We follow \citeil{Scott:2009}, and fix $\rho_{\mathrm{max}}=\langle\sigma v\rangle/(m_\u{DM}t_U)$, where $t_U$ is the age of the universe. For fixed $r_\u{tr}$, $\gamma$ and clump luminosity $r_p$ is fully determined, playing a role analogous to $\rho_0$ for the NFW and exponential profiles.

\section{\texorpdfstring{$e^\pm$}{e+/e-} from a clump}
\label{sec:cre}
We now discuss the calculation of the contribution of a nearby clump to the local electron-positron flux in the energies of interest for the DAMPE excess. Specifically, the next sub-section details the calculation of the flux of CRE from a clump, the following \cref{sec:clumpDAMPE} details the features of clumps that might explain the DAMPE observation, \cref{sec:linewidth} the constraints from the width of the DAMPE feature and, finally, \cref{sec:aniso} the constraints from the observed (lack of) anisotropy in the CRE arrival directions.

\subsection{Computing the \texorpdfstring{$e^\pm$}{e+/e-} spectrum}

Let $f_{e^\pm}^\u{clump}(\vec{x}, E) \equiv \du^4 N_{e^\pm}(\vec{x}, E)/\du E~\du^3x$ be the CRE distribution function due to DM annihilating in a clump. This distribution obeys the propagation equation \cite[see e.g.][]{Strong:2007nh}\footnote{We ignore convection and diffusive reacceleration since neither is relevant over the small volumes we are studying and the latter is not well-understood, especially at large distance from the Galactic center as those of interest here \cite{Orlando:2017tde}}.
\begin{align}
    \label{eq:prop_eq}
    \frac{\partial f_{e^\pm}^\u{clump}}{\partial t} &= D(E) ~ \nabla^2 f_{e^\pm}^\u{clump}(\vec{x}, E) + \frac{\partial}{\partial E} \left[ b(E) ~ f_{e^\pm}^\u{clump}(\vec{x}, E) \right] + Q^\u{clump}(\vec{x}, E).
\end{align}
The first term on the right hand side accounts for diffusion, which we assume to be homogeneous and isotropic. (However, see \citeil{Profumo:2018fmz} for a discussion of this point.) The energy-dependent coefficient, which we also assume to be homogeneous and isotropic, is parameterized as \cite{Orlando:2017tde}
\begin{align}
    D(E) &= D_0 \left( \frac{E}{E_0} \right)^\delta.
\end{align}
The standard values for these parameters are $D_0 = \SI{e28}{\centi\meter^2/\second}$, $E_0 = \SI{1}{\giga\electronvolt}$, and $\delta = 0.7$. Energy losses due to inverse-Compton scattering of cosmic microwave background and starlight photons as well as synchrotron emission are described by the second term of~\cref{eq:prop_eq}, with \cite{Colafrancesco:2005ji, Colafrancesco:2006he}
\begin{align}
    b(E) &= b_0 \left( \frac{E}{E_0} \right)^2.
\end{align}
We assume $b_0 \simeq \SI{e-16}{\giga\electronvolt/\second}$.

We will assume $f_{e^\pm}^\u{clump}$ is time-independent throughout so that the left-hand side of~\cref{eq:prop_eq} can be neglected. This stationarity assumption is justified when the distance the clump travels during the characteristic CR propagation time is much less than the clump's radius or distance from Earth $d$:
\begin{align}
    \label{eq:stationarity_condition}
    v_\u{clump}~t_\u{prop} \ll r_\u{clump},\ d.
\end{align}
The propagation timescale is $t_\u{prop} \sim \min(t_\u{diff}, t_\u{loss})$, where the diffusion and energy loss timescales are
\begin{align}
    \label{eq:prop_timescales}
    t_\u{diff} &= \frac{d^2}{D(E)},
    \qquad
    t_\u{loss} = \frac{E}{b(E)}.
\end{align}
The characteristic clump velocity is the Milky Way velocity dispersion $v_\u{clump} \sim \SI{220}{\kilo\meter/\second}$. At all the distances and scale radii considered in this work, we have checked that either:
\begin{enumerate}[(i)]
    \item the stationarity assumption holds;
    \item the clump is inconsistent with the narrow width of the DAMPE excess; or
    \item the clump is inconsistent with gamma ray constraints (see \cref{sec:gammas}).
\end{enumerate}
For concreteness, we assume the dark matter to be its own antiparticle. Then, for a DM clump centered at $\vec{x}_c$, the source term $Q^\u{clump}$ is
\begin{align}
    Q^\u{clump}(\vec{x}, E) &= \frac{\langle \sigma v \rangle}{2 m_\u{DM}^2} \rho^2(|\vec{x} - \vec{x}_c|) \frac{\du N_{e^\pm}}{\du E} (E).
\end{align}
In addition to the density profile, this quantity depends on the thermally-averaged product of the DM annihilation cross section and the relative velocity, $\langle \sigma v \rangle$; the DM mass $m_\u{DM}$; and the spectrum of $e^\pm$ per DM annihilation. We will only consider annihilations directly into $e^\pm$, since other final states will produce a wider excess than that observed by DAMPE, and require larger clump densities. The spectrum thus has the simple form
\begin{align}
    \frac{\du N_{e^\pm}}{\du E}(E) &= 2 \delta(E - m_\u{DM}).
\end{align}
\revised{We set the DM mass equal to $1513.6\us{GeV}$, the upper edge of the bin containing the excess. We comment on the impact of making $m_\u{DM}$ a free parameter in \cref{sec:linewidth}.}

\revised{The self-annihilation cross section is fixed to its thermal value, $\langle \sigma v \rangle = \SI{3e-26}{\centi\meter^3/\second}$. If the DM is not self-conjugate but is symmetric, $Q^\u{clump}(\vec{x}, E)$ must be multiplied by an additional factor of $1/2$. However, the thermal relic cross section is also increased by a factor of 2, which means our conclusions hold without modification.} Of course, it is possible to relax the assumption that the DM is thermally produced. Doing so would entail a trivial re-scaling of the clump luminosity and, in turn, of the normalization of the clump dark matter density.

The Green's function for the propagation equation (\cref{eq:prop_eq}) is known to be~\cite{Syrovatskii:1959}
\begin{align}
    G(\vec{x}, E; \vec{x}_s, E_s) &= \frac{1}{b(E)~[4 \pi \lambda(E, E_s)]^{3/2}} \exp \left[ -\frac{|\vec{x} - \vec{x}_s|^2}{4 \lambda(E, E_s)} \right],
\end{align}
where the energy loss and diffusion parameters appear in the combination
\begin{align}
    \lambda(E, E_s) &= \theta(E_s - E) \int_E^{E_s} \du E'\ \frac{D(E')}{b(E')}\\
                    &= \theta(E_s - E) \frac{D_0 E_0}{b_0 (1-\delta)} \left[ \left( \frac{E_0}{E} \right)^{1-\delta} - \left( \frac{E_0}{E_s} \right)^{1-\delta} \right].
\end{align}
We will place the Earth at the origin ($\vec{x} = \vec{0}$) and the clump's center at $\vec{x}_c = (0, 0, d)$. Performing the (trivial) integration over $E_s$ and using the clump's azimuthal symmetry, its contribution to the solution of the stationary propagation equation at Earth is
\begin{align}
    \label{eq:epm_distro}
    f_{e^\pm}^\u{clump} (E) &\equiv \int \du^3x_s\ \int \du E_s\ Q^\u{clump}(\vec{x}_s, E_s) ~ G(\vec{0}, E; \vec{x}_s, E_s)\\
                 &= \frac{2 \pi \langle \sigma v \rangle}{m_\u{DM}^2~b(E)~[4 \pi \lambda(E, m_\u{DM})]^{3/2}} \notag\\
                 &\hspace{1cm} \times \int_0^\infty \du r_s\ r_s^2 \exp\left[ -\frac{r_s^2}{4\lambda(E, m_\u{DM})} \right] \int_{-1}^1 \du c_{\theta_s}\ \left[ \rho\left( \sqrt{d^2 + r_s^2 - 2 d r_s c_{\theta_s}} \right) \right]^2.
\end{align}
The integral over $c_\theta$ can be performed analytically for the NFW and exponential profiles, but the radial integral must be evaluated numerically.

The CRE flux at Earth from the DM clump is thus
\begin{align}
    \phi_{e^\pm}^\u{clump} (E) &\equiv \frac{c}{4\pi} f_{e^\pm}^\u{clump}(E).
\end{align}

\subsection{Clumpy explanation for the DAMPE measurement}\label{sec:clumpDAMPE}

For fixed clump distance ($d$), radius ($r_s$ or $\rtidal$), and inner slope ($\gamma$), the \revised{density normalization ($\rho_0$) is} fixed through the requirement that the clump sources the DAMPE spike\revised{, which lies} in the energy bin $[1318.3,\, 1513.6]~\SI{}{\giga\electronvolt}$. \revised{Equating the integrated fluxes from the clump and background to the observed central value and labeling this bin with index $i_\u{excess}$ gives
\begin{align}
    \Phi_{e^\pm,i_\u{excess}}^\u{clump} + \Phi_{e^\pm,i_\u{excess}}^\u{bg} &= \Phi_{e^\pm,i_\u{excess}}^\u{obs} = \SI{9.75e-10}{\meter^{-2}s^{-1}\steradian^{-1}}.
\end{align}}
The second term is the \revised{integrated} CRE background \revised{flux}. \revised{For this w}e use the smoothly-broken power law fit by the DAMPE collaboration~\cite{Ambrosi:2017wek}:
\begin{equation}
    \label{eq:bg_dampe}
    \phi_{e^\pm}^\u{bg} = \phi_0 \left( \frac{\SI{100}{\giga\electronvolt}}{E} \right)^{\gamma_1} \left[ 1 + \left( \frac{E_b}{E} \right)^{(\gamma_1 - \gamma_2) / \delta} \right]^{-\delta},
\end{equation}
with parameter values
\begin{align}
    &\phi_0 = (1.62 \pm 0.01) \times \SI{e-4}{\meter^{-2}\second^{-1}\steradian^{-1}\giga\electronvolt^{-1}},\\\nonumber
    &E_b = 914 \pm \SI{98}{\giga\electronvolt},\\\nonumber
    &\gamma_1 = 3.09 \pm 0.01,\\\nonumber
    &\gamma_2 = 3.92 \pm 0.20.
\end{align}

\subsection{Line width constraint}
\label{sec:linewidth}

A clump sourcing the DAMPE spike must \revised{be narrow enough to avoid overproducing CREs in other energy bins. We set a constraint by requiring the integrated CRE flux in each energy bin (besides the one containing the excess)} to be within $3\sigma$ of the observed central value:
\revised{
\begin{align}
    \label{eq:lw_constraint}
    \max_i \frac{\Phi_\u{e^\pm,i}^\u{clump} + \Phi_\u{e^\pm,i}^\u{bg} - \Phi_\u{e^\pm,i}^\u{obs}}{\Delta \Phi_\u{e^\pm,i}^\u{obs}} &\leq 3,
\end{align}
}
where \revised{$i$ is the bin index and $\Delta \Phi_i$} is the upper error bar on the observed integrated flux.

\begin{figure}
    \centering
    \includegraphics[width=\textwidth]{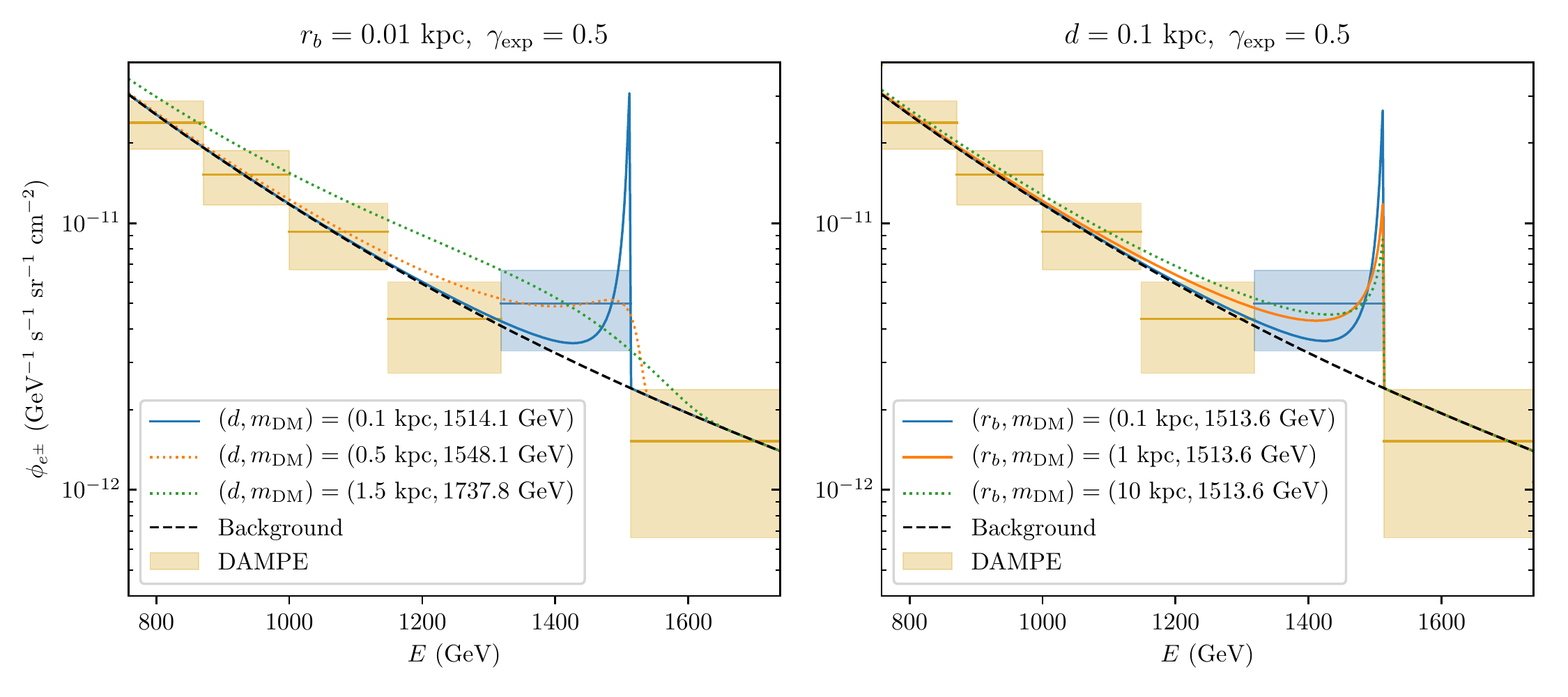}
    \caption{\revised{Differential CRE flux from exponential-profile clumps. The colored curves indicate the clump plus background flux as a function of distance (left panel) and scale radius (right). The dark matter mass was optimized over so as to minimize the line width constraint (left hand side of \cref{eq:lw_constraint}); the other clump parameters are fixed to the values indicated in the panels' titles. The black curve shows the background CRE flux. The DAMPE data with $\pm 3\sigma$ error bars is indicated by the colored boxes with the horizontal lines marking the central values. The bin containing the excess is highlighted in blue. The dotted curves are $3.4\sigma$, $7.2\sigma$ and $3.7\sigma$ above the flux in the bin below the excess. Ignoring this bin in the constraint calculation, the clump in the left panel with $d = \SI{1.5}{\kilo\parsec}$ is still $3.8\sigma$ larger than the flux in the second bin below the excess.}}
    \label{fig:lw_constraint}
\end{figure}

\revised{This constraint leads to the stringent requirement that the clump lies within about $\SI{0.4}{\kilo\parsec}$ of Earth, and also provides an upper bound on the scale radius for clumps with low inner slopes ($\gamma \sim 0.5$). To illustrate how the line width depends on the clump parameters, \cref{fig:lw_constraint} shows the CRE fluxes plus background for various clumps, along with the background prediction and DAMPE measurements. In the left panel, the clump distance is varied with the other parameters held fixed, while the scale radius is varied in the right panel.}

The resulting constraint can be understood by considering how the timescales for energy loss and diffusion (\cref{eq:prop_timescales}) vary with distance. For a $\sim\SI{1.5}{\tera\electronvolt}$ CRE these timescales are equal at $d \sim \SI{1}{\kilo\parsec}$. CREs produced in a \revised{clump further than roughly $\SI{1}{\kilo\parsec}$ from Earth} or \revised{in} the outer reaches of a nearby clump thus lose a substantial amount of energy, \revised{shifting CREs from the bin containing the excess into lower-energy bins.}

\revised{For illustration, in \cref{fig:lw_constraint}, we choose the dark matter mass that minimizes the left hand side of \cref{eq:lw_constraint}. The best-fit value ranges up to about $\SI{1700}{\tera\electronvolt}$. In principle, this optimization leads to a somewhat weaker line width constraint in comparison to the case of fixed DM mass. In practice, however, optimizing over the DM mass has only a marginal impact relative to our benchmark scenario $m_\u{DM} = \SI{1513.6}{\tera\electronvolt}$. This is because the CRE spectrum of a distant clump has a broad peak, as seen in the left panel of \cref{fig:lw_constraint}. As a result, increasing the DM mass shifts the spectrum to higher energies without dramatically reducing the CRE fluxes in lower-energy bins. A large increase in the DM mass leads to a CRE excess in the bin above the one containing the observed excess. Optimizing over $m_\u{DM}$ has a negligible impact on the other observational constraints considered herein, so we fix $m_\u{DM} = \SI{1513.6}{\giga\electronvolt}$ throughout the remainder of this work.}

\revised{It is also notable that the background CRE flux (\cref{eq:bg_dampe})} exceeds the value observed by DAMPE in the bin immediately below the one containing the excess ($[1148.2, ~ 1318.3]~\SI{}{\giga\electronvolt}$) by $\sim 2 \sigma$. \revised{The possibility that this bin is afflicted by some kind of systematic error has not been discussed in the literature. If this is the case, our line width constraint may be too strong. Excluding this bin when computing the line width constraint relaxes the bound on clump distance to roughly $\sim \SI{1.5}{\kilo\parsec}$. At larger distances, the CRE line from the clump is broad enough that it produces $3\sigma$ excesses in multiple bins, making this constraint very robust. But even if the line width constraint applies only to clumps beyond \SI{1.5}{\kilo\parsec}, our central conclusions about a connection between the DAMPE CRE feature and an enhancement to the local DM density still hold, as we will check in \cref{sec:results}.}

\subsection{Constraint from anisotropy}\label{sec:aniso}

Since the clump must be less than \SI{1}{\kilo\parsec} from Earth, it is important to determine whether it makes an observable contribution to the CRE dipolar anisotropy. The dipolar anisotropy is the magnitude $\delta_{e^\pm}$ of the first coefficient in the multipole expansion of the relative intensity~\cite{Strong:2007nh,Ahlers:2016rox}
\begin{align}
    \label{eq:anisodef}
    \frac{\phi_{e^\pm}(E, \theta, \phi)}{\phi_{e^\pm}^\u{tot}(E)} &\approx 1 + \vec{\delta}_{e^\pm}(E) \cdot \vec{n}(\theta, \phi),
\end{align}
where $\phi_{e^\pm}(E, \theta, \phi)$ and $\vec{n}(\theta, \phi)$ are the CRE flux and unit vector in the $(\theta, \phi)$ direction. In the diffusion approximation and neglecting heliospheric effects, the anisotropies for $e^\pm$ from an arbitrary DM clump and a point-like clump are respectively
\begin{align}
    \label{eq:aniso}
    \delta_{e^\pm}(E) &=
        \frac{3 D(E)}{c} \frac{\left| \nabla \phi_{e^\pm}^\u{clump}(E) \right|}{\phi_{e^\pm}^\u{total}(E)}\\
    &\approx
        \frac{3 D(E)}{c} \frac{2 d}{\lambda(E, m_\u{DM})} \frac{\phi_{e^\pm}^\u{clump}(E)}{\phi_{e^\pm}^\u{total}(E)},
\end{align}
where the gradient reduces to the derivative with respect to the clump's distance. By computing the bin-averaged anisotropy $\Delta_{e^\pm}$, this can be compared with experimental limits from Fermi~\cite{Abdollahi:2017kyf} and AMS-02~\cite{LaVacca:2016tqq}. The strongest constraint on the clumps we consider is Fermi's upper bound $\Delta_{e^\pm} < 2.33\ee{-2}$ over the energy range $[\SI{562}{\giga\electronvolt},\,\SI{1998}{\giga\electronvolt}]$.

\begin{figure}
    \centering
    \includegraphics{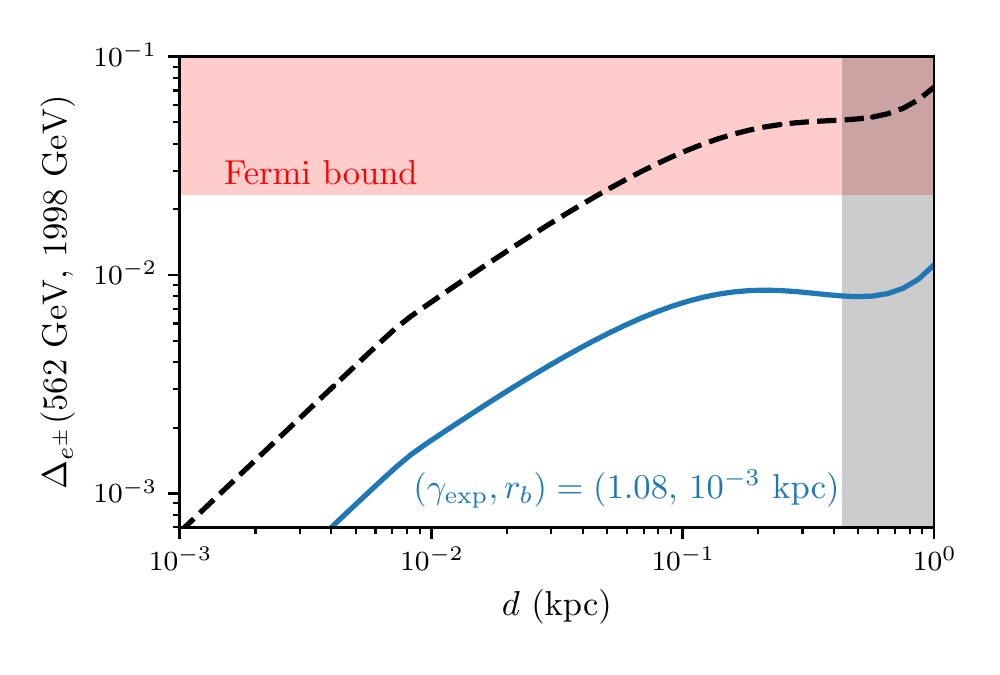}
    \caption{Bin-averaged anisotropies for a point-like (dashed black) and sharply-peaked exponential clump (blue) fit to the DAMPE excess. Fermi's upper limit over the indicated energy range is shown in red. In the grey region, the CRE spike from the point-like clump is too wide; the corresponding constraint for the exponential clump is very similar.}
    \label{fig:anisotropy}
\end{figure}

\Cref{fig:anisotropy} shows the anisotropy averaged over this bin for a point-like clump and sharply peaked exponential clump as a function of distance. Fermi's upper limit is displayed in red, and the grey region is the bound from requiring the CRE excess to be narrow (see previous section). For the point-like clump, the anisotropy constraint requires the clump to be nearly an order of magnitude closer to Earth than the line width constraint, which can be traced back to the extra factor of $d$ in \cref{eq:aniso}. However, even for the most point-like exponential-profile clump, the CRE anisotropy falls safely below the Fermi bound by a factor of a few, in agreement with another study of the DAMPE excess~\cite{Yuan:2017ysv}.

\section{Gamma rays from a clump}
\label{sec:gammas}

A clump is excluded by gamma ray observations if it is too bright as a point source or extended object. The expression for the differential gamma-ray flux is well-known:~
\begin{align}
    \label{eq:gamma_ray_flux}
    \phi_\gamma^\u{clump}(E, \theta) &= \frac{\Delta\Omega}{4\pi} \left[ \frac{\langle \sigma v \rangle}{2 m_\u{DM}^2} \frac{\du N_\gamma}{\du E}(E) \right] \left[ \frac{1}{\Delta\Omega} \int \du\Omega\ \int_0^\infty \du l\ \rho^2(l, \theta, \phi) \right].
\end{align}
In this work we only consider conical observing regions with radius $\theta$ centered on the clump, and use $\Delta\Omega$ to denote the corresponding solid angle. The first term in brackets depends on the particle physics properties of the DM, where $\du N_\gamma / \du E$ is the photon spectrum per annihilation. Since we assume the DM annihilates only into electrons, primary photons are produced through final state radiation (FSR). The electron mass is negligible compared to the center of mass energy for the annihilations, so the FSR spectrum factors from the hard process and the equivalent photon approximation applies~\cite{Peskin:1995ev}:
\begin{align}
    \frac{\du N_\gamma}{\du E} &= \frac{\alpha_\u{EM}}{\pi m_\u{DM}} \frac{1 + (1-x)^2}{x} \left[ -1 + \log\left( \frac{4 (1-x) m_\u{DM}^2}{m_e^2} \right) \right],
\end{align}
where $x \equiv E / m_\u{DM}$. The second bracketed term in \cref{eq:gamma_ray_flux} is the $J$-factor, where the integral is performed along the line of sight. For all clumps considered in this work, the clump's gamma ray flux is orders of magnitude larger than the gamma ray flux from DM annihilating in the galactic center.

\begin{figure}
    \centering
    \includegraphics[width=\textwidth]{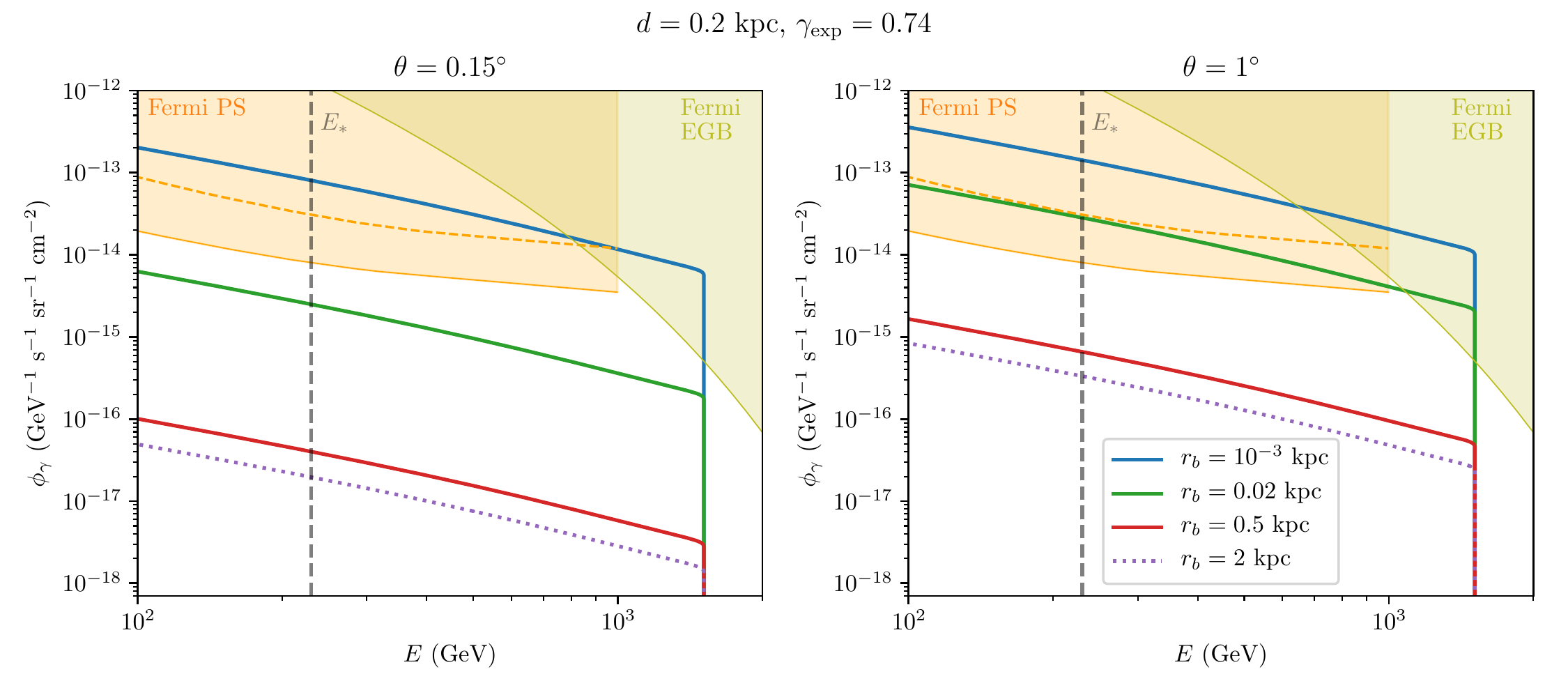}
    \caption{Gamma ray fluxes from DM clumps with the same exponential profiles as in \cref{fig:lw_constraint}, with conical observing regions whose angular radii are indicated by the subplot titles. The shaded areas show Fermi-LAT's point source sensitivity and the measured extragalactic background. The dashed orange line is the sensitivity to a source near $(b, l) = (0^\circ, 0^\circ)$ while the lower boundary of the orange region is the sensitivity to a source at $(120^\circ, 45^\circ)$. \revised{The vertical line at $E_* = \SI{230}{\giga\electronvolt}$ is the energy at which the clump would be observed as a point source most easily.}}
    \label{fig:gamma_spectra_constraints_exp}
\end{figure}

\Cref{fig:gamma_spectra_constraints_exp} exhibits the gamma ray spectra for exponential clumps with various scale radii. The subplots are for observing regions with radii $\theta_\u{PSF} \equiv 0.15^\circ$ (roughly Fermi's angular resolution above \SI{100}{\giga\electronvolt}~\cite{Atwood:2009ez}) and $1^\circ$. The solid line bounding the orange region from below shows $\phi_\gamma^\u{PS}(E)$, Fermi's sensitivity to point sources with an arbitrary power law spectrum~\cite{FermiInfo}. \revised{Since the sensitivity curves are provided for a limited set of source positions by the Fermi Collaboration, we optimistically set the clump's location to $(b, l) = (120^\circ, 45^\circ)$ to obtain the strongest constraint. However, for a clump aligned with the Galactic Center ($(b, l) = (0^\circ, 0^\circ)$) the constraint is only a factor of $\sim 5$ weaker, as shown by the dashed orange line.} Since Fermi has not detected point-like emission from DM subhalos\footnote{
Note, however, that two objects in the Fermi catalogue exhibit statistically significant spatial extension~\cite{Xia:2016uog}. The lack of \revised{corresponding signals} in other wavelengths suggests this could arise from two point sources that cannot be individually resolved by Fermi~\cite{Chou:2017wrw}.
}, the emission in a $\theta_\u{PSF}$ cone around a DM clump may not exceed this minimum flux over Fermi's energy range. The constraint on a clump treated as a point source is thus
\begin{align}
    \label{eq:gammapsconstraint}
    \phi_\gamma^\u{clump}(E, \theta_\u{PSF}) &< \phi_\gamma^\u{PS}(E), \qquad \SI{30}{\mega\electronvolt} \lesssim E \lesssim \SI{1}{\tera\electronvolt}.
\end{align}
As can be seen from the left panel of \cref{fig:gamma_spectra_constraints_exp}, the clump with $\rtidal=\SI{0.001}{\kilo\parsec}$ is excluded on these grounds. 

\revised{The gamma-ray energy $E_*$ at which the clump could be observed as a point source is the value for which the above inequality is closest to being violated. Since we have fixed the dark matter mass, this energy is also fixed: $E_* = \SI{230}{\giga\electronvolt}$. This is indicated by the vertical dashed line in the figure, and will be used as a reference gamma-ray energy in our final constraint plots.}

The significant difference in fluxes between the $\theta_\u{PSF}$ and $1^\circ$ observing region shows the clump's gamma ray emission is spatially extended over much of the parameter space we consider. To estimate the constraint on extended emission from the clump, we compare the flux in a $1^\circ$ cone to Fermi's extragalactic background (EGB) model, conservatively requiring the former to be no more than an order of magnitude larger than the later:
\begin{align}
    \label{eq:gammaextendedsrcconstraint}
    \phi_\gamma^\u{clump}(E, 1^\circ) &< 10 ~ \phi_\gamma^\u{EGB}(E).
\end{align}
For the EGB flux, we use Model A from~\citeil{Ackermann:2014usa}:
\begin{equation}
    \phi_\gamma^\u{EGB}(E) = I_{100} \left( \frac{E}{\SI{0.1}{\giga\electronvolt}} \right)^{-\gamma} e^{-E / E_\u{cut}},
    \qquad
    \begin{cases}
    I_{100} = \SI{1.48e-7}{\giga\electronvolt^{-1}\centi\meter^{-2}\second^{-1}\steradian^{-1}},\\
    \gamma = 2.31,\\
    E_\u{cut} = \SI{362}{\giga\electronvolt}.
    \end{cases}
\end{equation}
This is illustrated by the yellow shaded region in \cref{fig:gamma_spectra_constraints_exp}.

We use a factor of 10 in \cref{eq:gammaextendedsrcconstraint} because two other important extended sources, the Fermi bubbles~\cite{Su:2010qj} and Cygnus cocoon~\cite{Grenier:2012rta}, exceed the EGB by \revised{approximately} the same amount. \revised{Another bright, extended source is the Galactic plane, which is a factor 10-100 brighter than the isotropic extragalactic background, depending on Galactic latitude~\cite{Abramowski:2014vox}. Should the clump be located near an extended source a factor of $R$ brighter than the EGB, an increase in observation time by roughly a factor of $\sqrt{R}$ would be needed to achieve a comparable sensitivity to what we have assumed~\cite{Ackermann:2014usa}.}

The same $1^\circ$-sized observing region was used in a different study of the DAMPE excess~\cite{Yuan:2017ysv}, and is comparable to the extent of the Cygnus cocoon. We ignore galactic diffuse emission, as it is subdominant near $\SI{1}{\tera\electronvolt}$. Due to the shape of the clump's gamma-ray spectrum, the constraints from treating it as a point source and an extended source are quite similar.

We note that, as always, the production of high-energy electrons and positrons yields secondary emission from energy loss processes~\cite{Colafrancesco:2005ji, Colafrancesco:2006he,Jeltema:2008hf,Storm:2012ty,McDaniel:2017ppt,Cirelli:2009vg, Profumo:2009uf,Jeltema:2011bd,McDaniel:2018vam}. Here we note the low magnetic fields and background radiation density in conjunction with the diffusion of high-energy electrons and positrons make the synchrotron and inverse Compton emission both very diffuse and tenuous. An estimate of the secondary emission would thus require a detailed model of the Galactic clump distribution over large distance scales, implying large systematic uncertainties---for instance, is the clump responsible for DAMPE an unlikely nearby over-density, or is it part of a clumpy dark matter density distribution? Even with some of the assumptions we do make in \cref{sec:clumpstats} to calculate the probability of finding a clump of given size and distance would not eliminate this systematic uncertainty. Additionally, while irrelevant for our discussion here, calculating the secondary emission would also introduce the systematic uncertainty related to the density profile of the smooth dark matter component.

If one were to calculate the secondary emission from high-energy CREs from the clump alone, the resulting gamma-ray flux would always be significantly more extended than the flux of gamma rays from prompt radiative emission as calculated in this section. Since the prompt emission is already considerably extended, the secondary signal would be spread over large angular regions, and would consist of photons with such low energies that the galactic background would present an obstacle to obtaining constraints. A rough estimate of the angular extension of the secondary emission is as follows: consider CRE in the DAMPE energy range, $E_{e^\pm}\simeq \SI{1.3}{\tera\electronvolt}$. The relevant timescale for secondary emission is $\tau\sim \tau_{\rm loss}\sim E/b(E)\simeq 8\times \SI{e12}{\second}$, and the associated spatial extent of the CRE is $d\sim \sqrt{D(E)\tau}\sim \SI{1}{\kilo\parsec}$. \revised{But we argued in \cref{sec:linewidth} that the distance to the clump is at most $\sim\SI{1}{\kilo\parsec}$, and we will confirm this directly in when we present our full results in \cref{sec:results}. Thus,} the angular extent of the secondary emission is always much larger than several 10s of degrees.

\section{Local clump abundance}\label{sec:clumpstats}
Now we study the statistical considerations for finding a nearby dark matter clump to account for the DAMPE excess. In principle, a clump of arbitrarily high luminosity might be found within any given distance from the observer. However, simulations of substructure formation give quantitative predictions of the abundance and clustering of such objects, so therefore, we can estimate the probability that a clump satisfying some set of criteria lies within a given distance of the observer. In this section, we describe the calculation of such probabilities from $N$-body simulation outputs.

\subsection{Clump criteria and assumptions}
\label{sec:likelihood-criteria}
The primary criterion of interest for local dark matter clumps is their detectability in either the gamma-ray or cosmic-ray spectra. The flux of particles of either kind originating from a clump is dependent on two quantities: the distance $d$ between the observer and the clump, and the luminosity of the clump, defined as
\begin{equation}
    \label{eq:luminosity}
    L = \frac{\left\langle\sigma v\right\rangle}{2m_\u{DM}^2} {\int_{\mathrm{clump}} \du^3 \bm{\mathrm{r}}} \, \rho(\bm{\mathrm{r}})^2.
\end{equation}
In this work, we are interested in very nearby clumps ($d\lesssim\SI{1}{\kilo\parsec}$), so an additional criterion becomes relevant: the radius $r$ of the clump itself. Extended clumps produce a more diffuse signal, and the associated broadening of the spectrum can rule out such clumps as the origin of a narrow excess, as explained in~\cref{sec:linewidth}. Thus, we ask: \emph{for fixed $r_{\mathrm{max}}$, $d_{\mathrm{max}}$, and $L_{\mathrm{min}}$, what is the probability that a halo with radius $r<r_{\mathrm{max}}$ and luminosity $L>L_{\mathrm{min}}$ is found within a distance $d<d_{\mathrm{max}}$?} This question can be answered using clump parameter distributions drawn from $N$-body simulations.

Note that $L_{\mathrm{min}}$ is itself dependent on $d_{\mathrm{max}}$: the further away a clump is, the greater the luminosity required in order for it to be observed. It is important for our purposes that $L_{\mathrm{min}}$ is a strictly increasing function of $d_{\mathrm{max}}$, since this means that a clump with $L>L_{\mathrm{min}}(d_{\mathrm{max}})$ is observable no matter where it lies within a radius $d_{\mathrm{max}}$. Less obvious is the fact that $L_{\mathrm{min}}$ is also an increasing function of $r_{\mathrm{max}}$, as shown in the top-left panels of~\cref{fig:contours_nfw_0-5,fig:contours_nfw_1,fig:contours_exp_0-52,fig:contours_exp_0-74,fig:contours_exp_1-08}. Thus, similarly, halos with $r<r_{\mathrm{max}}$ are the ones that contribute to indirect detection observables, even as they contribute less to direct detection. We stress that since we impose $r<r_{\mathrm{max}}$, the likelihoods we calculate  cannot be used to assess the probability of having a particular \emph{terrestrial} overdensity of dark matter. They pertain only to the quantities associated with indirect detection. The full probabilistic connection between line-like indirect detection features and direct detection observables will be explored in a forthcoming study~\citep{WorkInPrep}.

In calculating the probability that a clump with given properties is observed within a given distance, we make the assumption that clumps have a uniform random spatial distribution. This is clearly a poor assumption for $d\sim\SI{10}{\kilo\parsec}$, but our region of interest is for $d\lesssim\SI{1}{\kilo\parsec}$ at most. At such short distances, clustering does not have a significant effect on the spatial distribution of clumps. With this assumption, it suffices to find the number density $n_s$ of clumps that meet the criteria of interest. Once $n_s$ is found, the likelihood of finding such an object within a distance $d_{\mathrm{max}}$ is just the probability that an observer placed into a random field of points with this number density happens to lie within $d_{\mathrm{max}}$ of at least one. This probability is given by $1-\exp(-n_sV)$, where $V$ is the volume of a ball of radius $d_{\mathrm{max}}$.

An important uncertainty in this computation comes from the mass function of nearby dark matter clumps. While we draw the mass function from $N$-body simulations~\citep{Garrison-Kimmel:2014}, this neglects potential baryonic effects on the population of subhalos in the galactic disk. Simulations suggest that the galactic disk destroys smaller clumps, depleting the mass function by a factor of 2--5 for masses below $\sim10^9M_\odot$ at our distance from the galactic center \citep{Kelley:2018,Richings:2018}. We do not include this effect in our calculation of probabilities, but since the sole impact would be a uniform and linear reduction in the likelihoods, this does not substantially alter our conclusions.

\subsection{Calculation of number densities}
\label{sec:number-densities}
Schematically, the number density $n_s$ of clumps satisfying our criteria can be found from the overall number density $n_0$ of clumps along with relevant parameter distributions. Suppose that the density profile of a clump is described by a set of parameters $\bb\alpha=\{\alpha_1,\dotsc,\alpha_N\}$, with a joint cdf $P(\bb\alpha)$, and suppose that the criteria are described by parameters $\bb\beta=\{\beta_1,\dotsc,\beta_M\}$. We can then represent our criteria by the function
\begin{equation}
	\chi(\bb\alpha;\,\bb\beta)=\begin{cases}
		1 & \text{the clump }\bb\alpha\text{ satisfies the criteria }\bb\beta\\
		0 & \text{otherwise.}
	\end{cases}
\end{equation}
Then, in general, the number density of such objects is given by
\begin{equation}
n_s(\bb\beta)=n_0\int\du^N\bb\alpha\,\frac{\du^NP(\bb\alpha)}{\du\alpha_1\dotsm\du\alpha_N}\chi(\bb\alpha;\,\bb\beta).
\end{equation}
Now the problem is reduced to specifying the indicator function $\chi$ and the joint probability distribution $P$. Both of these depend on the clump density profile.

As discussed in \cref{sec:dmclumps}, a standard NFW halo is described by two parameters, typically the density normalization $\rho_0$ and the scale radius $r_s$, or the concentration $c$ and the virial mass $M_{\mathrm{vir}}$. The distribution of concentration at fixed mass has been studied extensively in $N$-body simulations, so it is convenient to parametrize NFW halos in terms of $c$ and $M_{\mathrm{vir}}$. The generalized NFW profile that we employ in this work has an additional dependence on a parameter $\gamma$, which we fix for simplicity. Then the number density $n_s$ takes the form
\begin{equation}
    \label{eq:probability-nfw}
    n_s(d_{\mathrm{max}},r_{\mathrm{max}})=n_0\int\du M_{\mathrm{vir}}\,\du c\,\frac{\du P(M_{\mathrm{vir}})}{\du M_{\mathrm{vir}}}\frac{\du P(c\,|\,M_{\mathrm{vir}})}{\du c}\chi_{\mathrm{NFW}}(c,M_{\mathrm{vir}};\,d_{\mathrm{max}},r_{\mathrm{max}}).
\end{equation}
The indicator function $\chi_{\mathrm{NFW}}$ is 1 when the specified $c$, $M_{\mathrm{vir}}$, and $\gamma$ describe a halo with $r_s<r_{\mathrm{max}}$ and $L>L_{\mathrm{min}}(d_{\mathrm{max}},r_{\mathrm{max}})$.

The distribution of clump masses, $\du P(M_{\mathrm{vir}})/\du M_{\mathrm{vir}}$, is drawn from the mass function of \citeil{Hooper:2016cld}, namely
\begin{equation}
\label{eq:mass-function}
\frac{\du N}{\du M\,\du V}=\frac{628}{M_\odot\SI{}{\kilo\parsec^3}}\left(\frac{M}{M_{\odot}}\right)^{-1.9}.
\end{equation}
The distribution $\du P(c\,|\,M_{\mathrm{vir}})/\du c$ can be computed from a concentration-mass relation. For standard NFW ($\gamma=1$), we use the relation of \citeil{Sanchez-Conde:2014}. However, for $\gamma=0.5$, the concentration-mass relation has a different form. Fits to simulated halos exhibit a degeneracy between $\gamma$ and $c$: fixing $\gamma$ at a smaller value leads to a larger best-fit value of $c$ \citep{Ricotti:2007}. Since the concentration-mass relation has not been extensively studied as a function of $\gamma$, it is difficult to precisely characterize the impact on our calculation. Following \citeil{Ricotti:2007}, we suppose that at fixed mass, a profile fit with $\gamma=0.5$ will have $c$ increased by a factor of $\sim1.5$ compared with the value expected from \citeil{Sanchez-Conde:2014}. However, we emphasize that this is a significant uncertainty in computing the abundance of halos above a luminosity threshold when $\gamma\neq1$.

The exponential (tidally-truncated profile) is described by parameters $\rtidal$, $\rho_0$, and $\gamma$. We choose fixed values of $\gamma$, as with the parameter $\gamma$ in the NFW case. We use the distribution $\du P(\rtidal\,|\,M_{\mathrm{sub}})/\du \rtidal$ given in \citeil{Hooper:2016cld}, and we use the mass function of~\cref{eq:mass-function}. Then the number density has the form
\begin{equation}
    \label{eq:probability-exp}
    n_s(d_{\mathrm{max}},r_{\mathrm{max}})=n_0\int\du M_{\mathrm{sub}}\,\du \rtidal\,\frac{\du P(M_{\mathrm{sub}})}{\du M_{\mathrm{sub}}}\frac{\du P(\rtidal\,|\,M_{\mathrm{sub}})}{\du M_{\mathrm{sub}}}\chi_{\mathrm{exp}}(M,\rtidal;\,d_{\mathrm{max}},r_{\mathrm{max}}).
\end{equation}
As in the NFW case, the indicator function $\chi_{\mathrm{NFW}}$ is 1 when the specified $\rtidal$, $M$, and $\gamma$ describe a halo with $\rtidal<r_{\mathrm{max}}$ and $L>L_{\mathrm{min}}(d_{\mathrm{max}},r_{\mathrm{max}})$.

We evaluate the integrals of~\cref{eq:probability-nfw,eq:probability-exp} numerically. The resulting probabilities are shown in terms of the corresponding $z$-scores (sigmas) in the $(d_{\mathrm{max}},r_{\mathrm{max}})$ plane in the bottom-right panels of~\cref{fig:contours_nfw_0-5,fig:contours_nfw_1,fig:contours_exp_0-52,fig:contours_exp_0-74,fig:contours_exp_1-08}. In each case, the behavior of the likelihoods as a function of distance and halo radius is nearly the same. The threshold luminosity is only weakly dependent on the halo radius, so at fixed distance, the likelihood is constant as long as the halo is massive enough to overcome the luminosity threshold. Since the halo radius is related to its luminosity via the fairly narrow concentration-mass relation, this produces a sharp cutoff in the likelihood that evolves with the distance. On the other hand, at fixed halo radius, the likelihood at first increases approximately linearly with distance due to the larger available volume ($\sim d^3$), despite the increase in the luminosity threshold ($\sim d^2$). However, for $d\gtrsim\SI{1}{\kilo\parsec}$, diffusion effects are so strong that the luminosity threshold increases exponentially, and the likelihood falls off sharply again. As a result, the most probable region includes clumps with radius $\sim\SI{1}{\kilo\parsec}$ at a distance $\sim\SI{1}{\kilo\parsec}$.

Since $\gamma$ is held fixed in~\cref{fig:contours_nfw_0-5,fig:contours_nfw_1,fig:contours_exp_0-52,fig:contours_exp_0-74,fig:contours_exp_1-08}, the likelihoods therein should be interpreted as relative probabilities. Realistically, $\gamma$ will follow some distribution, and the true probability is found by integrating over this distribution. For instance, in the exponential case, \citeil{Hooper:2016cld} find that the distribution of $\gamma$ is approximately independent of mass, with the form
\begin{equation}
\label{eq:gamma-distribution}
    \frac{\du P(\gamma)}{\du\gamma} = 
        \frac{1}{\sqrt{2\pi}}\frac{1}{
            \sigma-\kappa(\gamma-\langle\gamma\rangle)
        } \exp\left(
            -\frac{\log^2\left[
                1-\kappa(\gamma-\langle\gamma\rangle)/\sigma
            \right]}{2\kappa^2}
        \right),
\end{equation}
and with parameter values $\langle\gamma\rangle=0.74$, $\kappa=0.10$, and $\sigma=0.42$. Thus, the number density should be calculated as 
\begin{equation}
    \label{eq:probability-exp-gamma}
    n_s(d_{\mathrm{max}},r_{\mathrm{max}})=n_0\int\du M\,\du \rtidal\,\du\gamma\,\frac{\du P(M)}{\du M}\frac{\du P(\rtidal\,|\,M)}{\du M}\frac{\du P(\gamma)}{\du\gamma}\chi_{\mathrm{exp}}(M,\rtidal,\gamma;\,d_{\mathrm{max}},r_{\mathrm{max}}).
\end{equation}
\revised{In the following section, we evaluate the corresponding likelihoods and compare with those likelihoods calculated at fixed $\gamma$.}

\section{Results and Discussion}\label{sec:results}

\begin{figure}
    \centering
    \includegraphics[height=540pt]{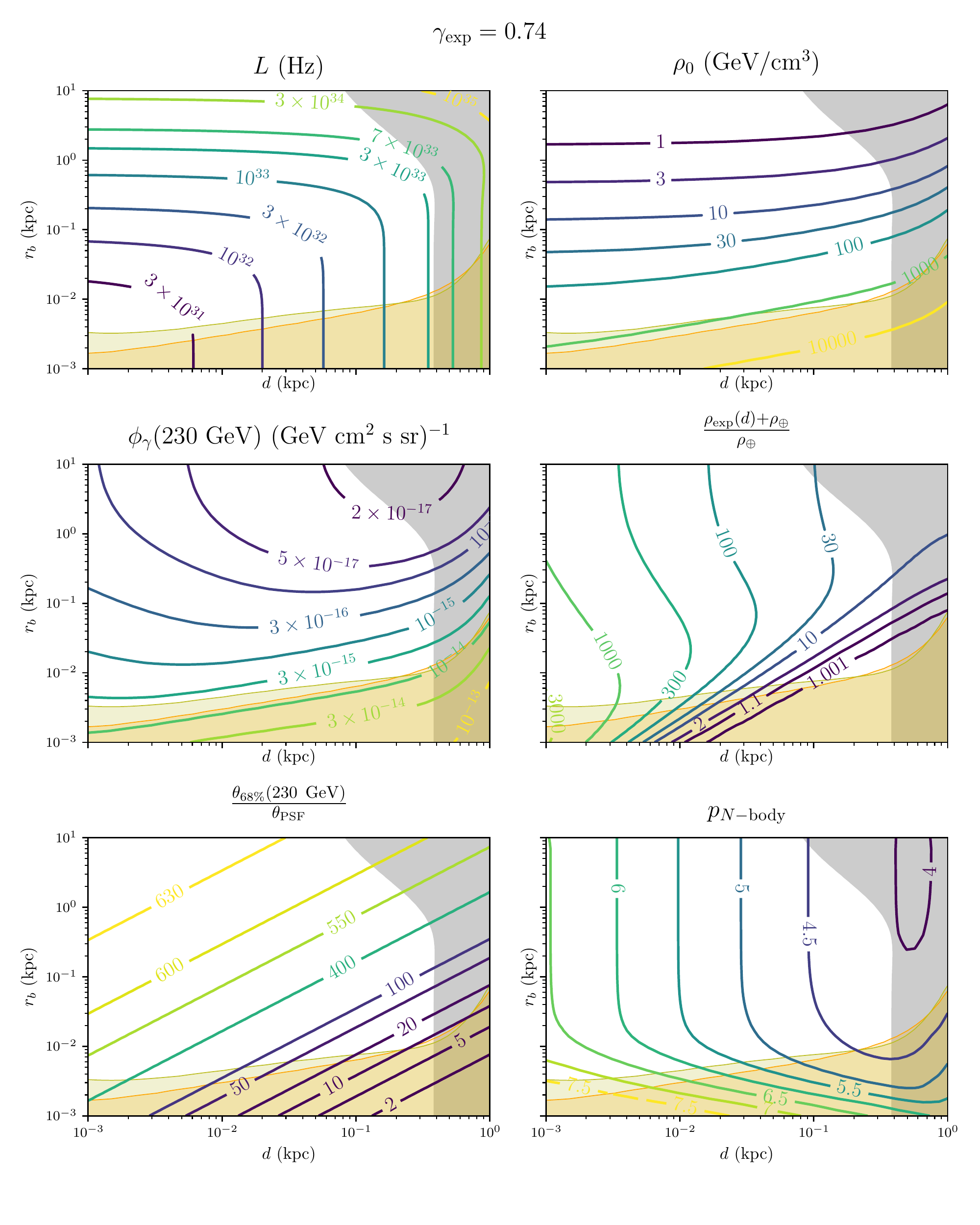}
    % Shift the caption up so that LaTeX knows this figure fits on one page
    \vspace{-1cm}
    \caption{Characteristics of a exponential clump with $\gamma=0.74$ capable of sourcing the DAMPE excess as a function of its distance from Earth $d$ and scale radius. The subplots show the clump's luminosity, density normalization, gamma-ray spectrum at $E_\gamma = \SI{230}{\giga\electronvolt}$ in a $0.15^\circ$ observing region, relative contribution to the local DM density, the gamma-ray emission's spatial extent at $E_\gamma = \SI{230}{\giga\electronvolt}$, and the $N$-body probability ($z$-score) of finding a clump at least as compact and luminous. The gray region is ruled out since the resulting $e^\pm$ line would be too wide. The orange and yellow regions are excluded by Fermi constraints on point and extended sources, respectively. See the text for more details.}
    \label{fig:contours_exp_0-74}
\end{figure}

\begin{figure}[p]
    \centering
    \includegraphics[height=540pt]{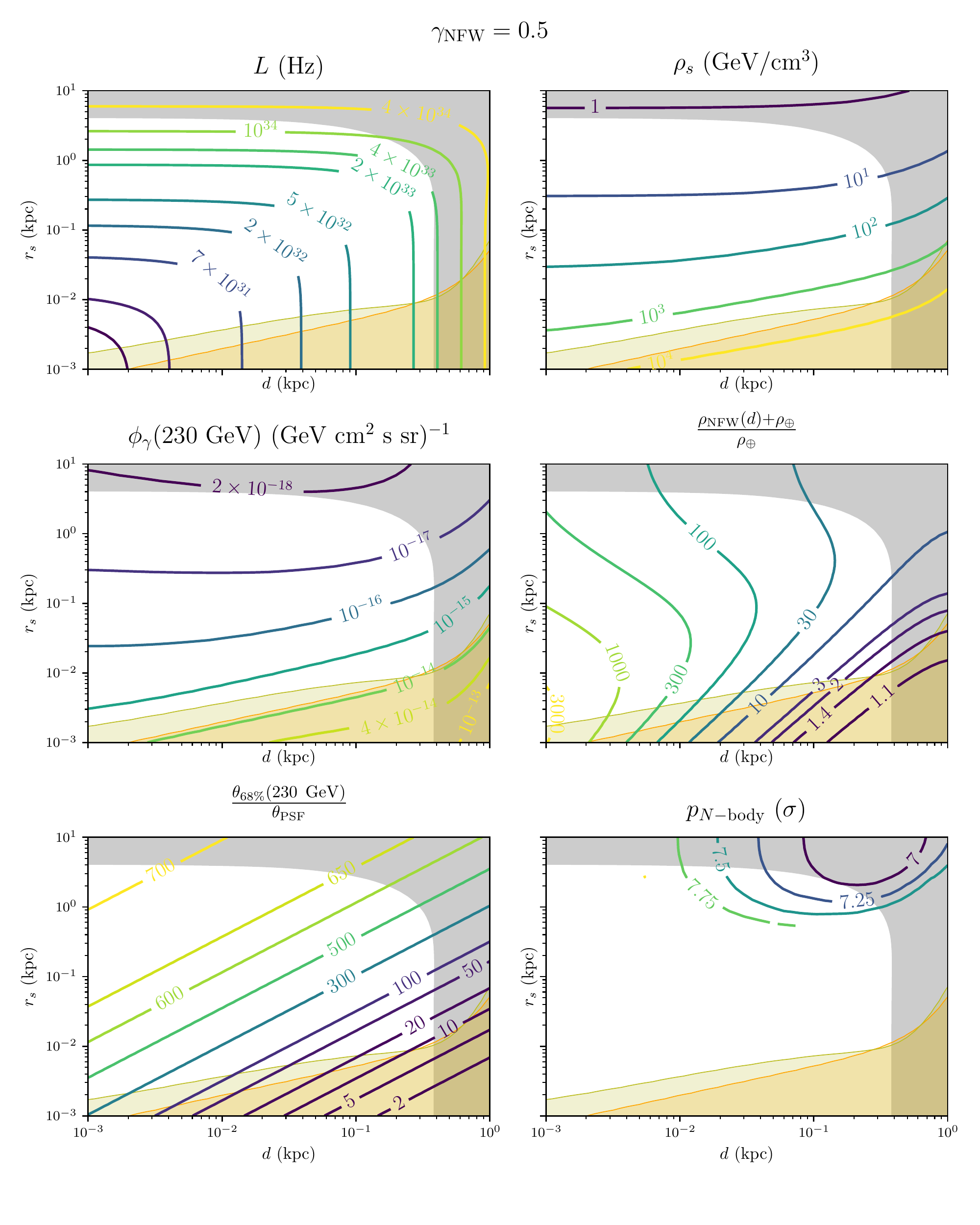}
    \caption{As in~\cref{fig:contours_exp_0-74}, but for an NFW clump with $\gamma=\frac12$.}
    \label{fig:contours_nfw_0-5}
\end{figure}

\begin{figure}[p]
    \centering
    \includegraphics[height=540pt]{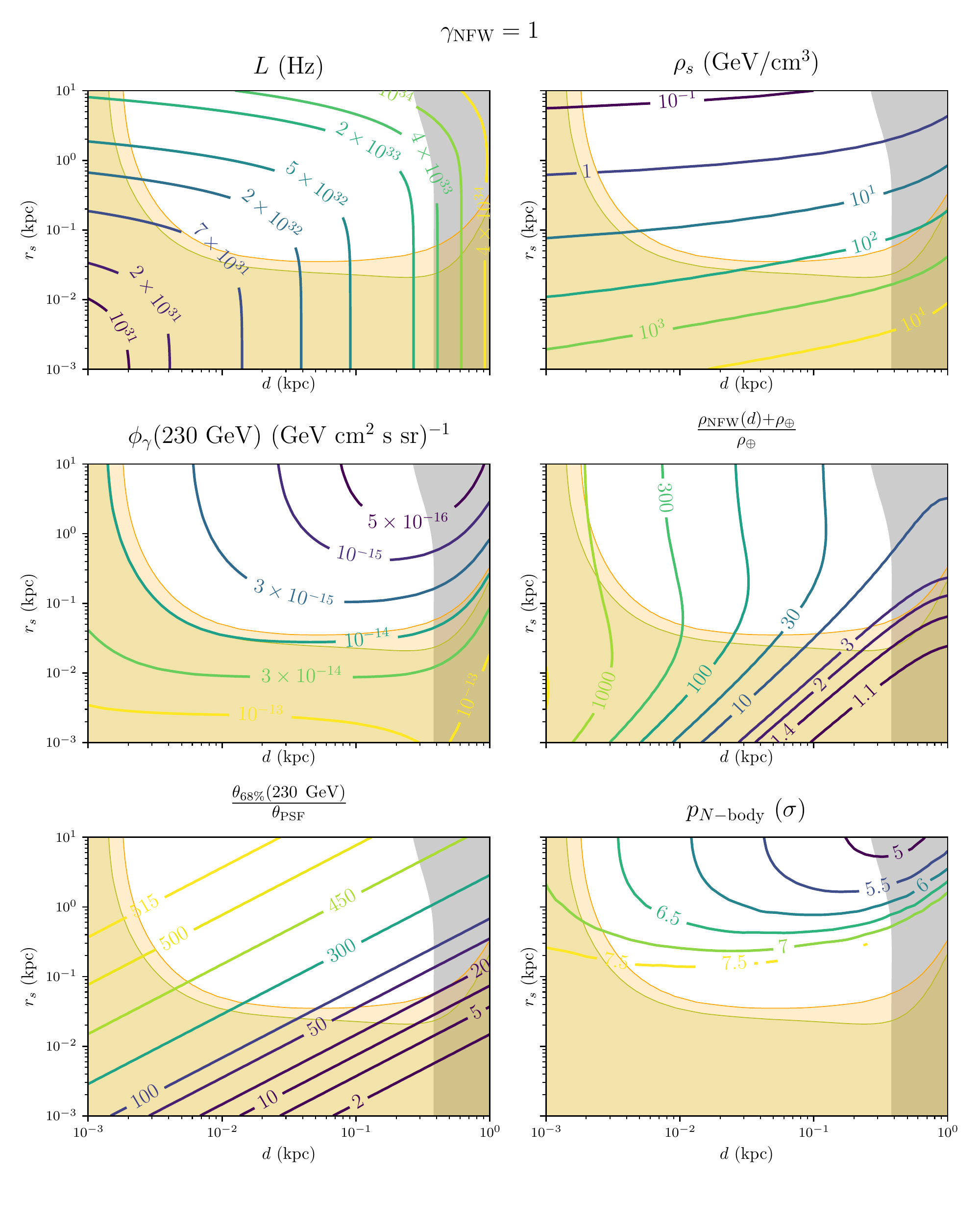}
    \caption{As in \cref{fig:contours_exp_0-74}, but for an NFW clump with $\gamma = 1$.}
    \label{fig:contours_nfw_1}
\end{figure}

\begin{figure}[p]
    \centering
    \includegraphics[height=540pt]{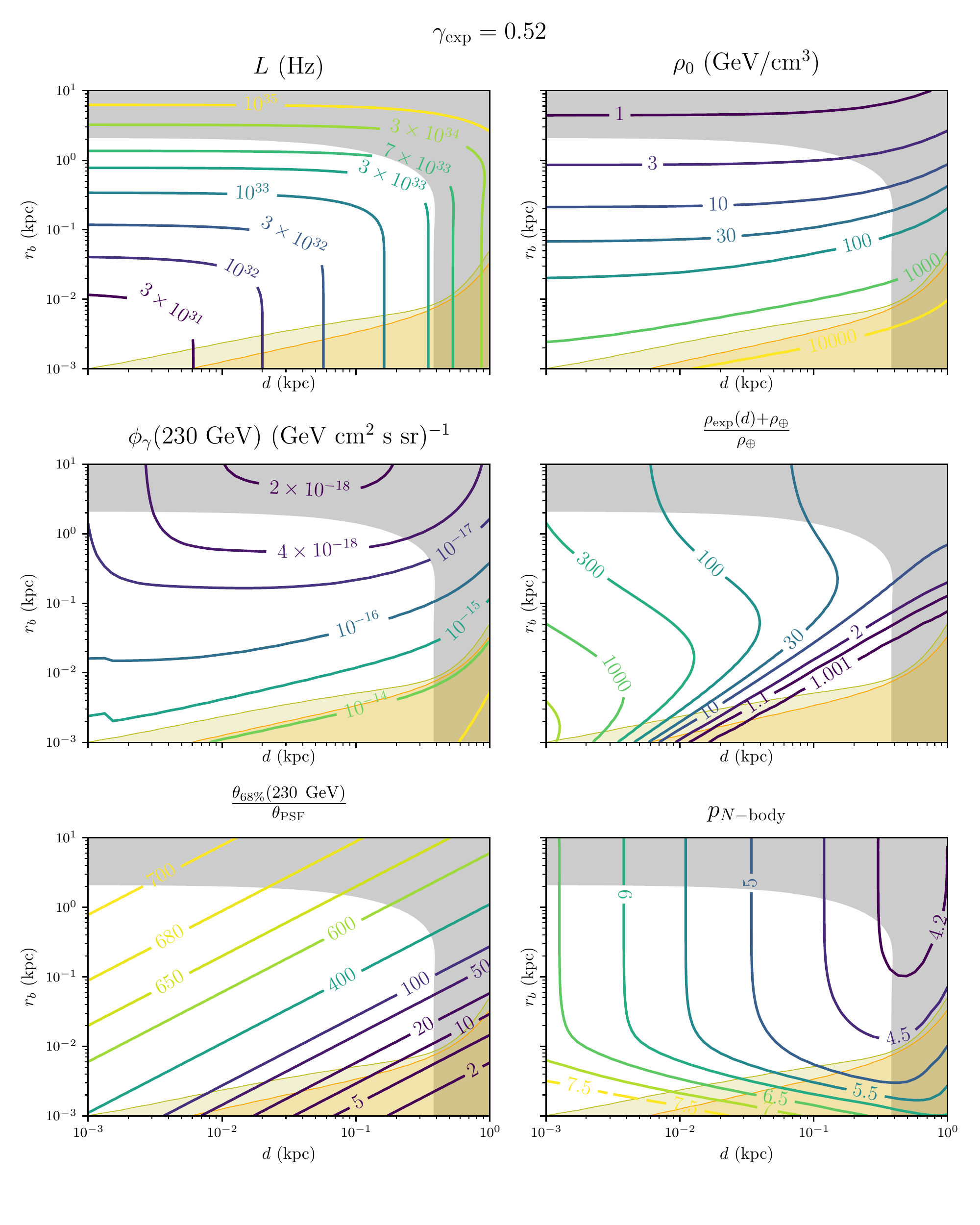}
    \caption{As in \cref{fig:contours_exp_0-74}, but for an exponential clump with $\gamma = 0.52$.}
    \label{fig:contours_exp_0-52}
\end{figure}

\begin{figure}[p]
    \centering
    \includegraphics[height=540pt]{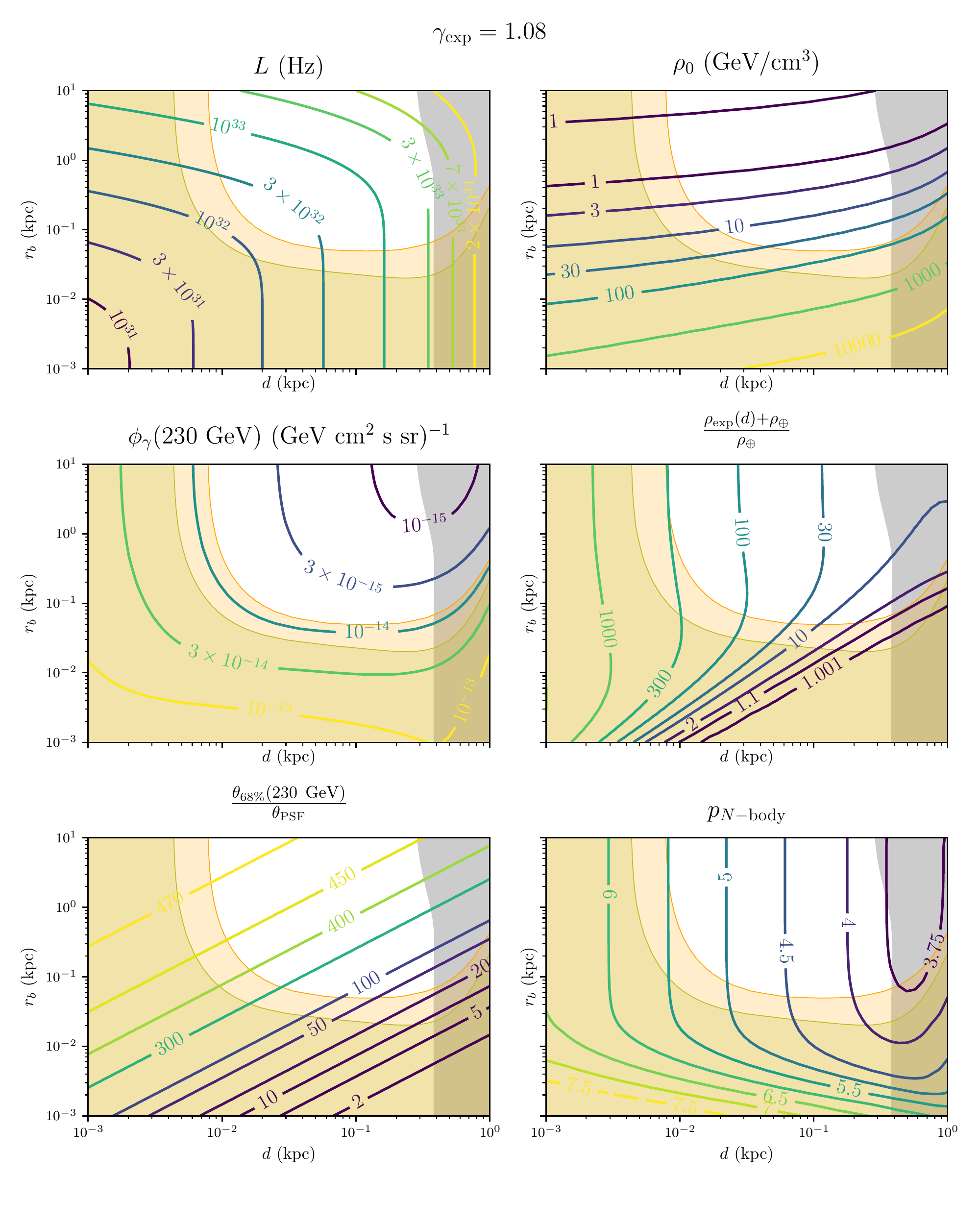}
    \caption{As in \cref{fig:contours_exp_0-74}, but for an exponential clump with $\gamma = 1.08$.}
    \label{fig:contours_exp_1-08}
\end{figure}

The characteristics of a clump required to explain the DAMPE excess are collected in \cref{fig:contours_nfw_0-5,fig:contours_nfw_1,fig:contours_exp_0-52,fig:contours_exp_0-74,fig:contours_exp_1-08}. \Cref{fig:contours_exp_0-74} shows results for a typical exponential clump with an  inner power law index of $\gamma=0.74$. The same results for different parameter values and for an NFW profile are shown after the text in~\cref{fig:contours_nfw_0-5,fig:contours_nfw_1,fig:contours_exp_0-52,fig:contours_exp_1-08}. The contours in each subplot show:
\begin{itemize}
    \item Luminosity $L$ (\cref{eq:luminosity}), top-left panel;
    \item The DM profile's density normalization $\rho_0$, top-right panel;
    \item The gamma-ray flux at $\SI{230}{\giga\electronvolt}$ in a $0.15^\circ$ observing cone (see discussion following \cref{eq:gammapsconstraint}), center-left panel;
    \item The local DM density enhancement from the clump, center-right panel;
    \item A measure of the clump's spatial extent in gamma-rays, $\theta_{68\%}$, the angular radius of the cone containing $68\%$ of the clump's gamma-ray flux at $E_\gamma = \SI{230}{\giga\electronvolt}$~\cite{Chou:2017wrw}, bottom-left panel;
    \item The $N$-body probabilities (discussed in the previous section), bottom-right panel, expressed in number of $\sigma$, bottom-right panel.
\end{itemize}
The shaded regions are excluded by:
\begin{itemize}
    \item Fermi's gamma-ray point source constraint (\cref{eq:gammapsconstraint}, orange region);
    \item Constraints on extended gamma-ray emission (\cref{eq:gammaextendedsrcconstraint}, yellow region);
    \item The CRE excess' width (\cref{sec:linewidth}, grey region).
\end{itemize}

In general, we find the line width constraint forces the clump to lie within $\sim \SI{0.3}{\kilo\parsec}$ of Earth, and also provides an upper bound on the radius for clumps with small inner slope. The gamma-ray constraints give a complementary lower bound on the radius and a minimum distance from Earth of $\sim 10^{-3} - \SI{e-2}{\kilo\parsec}$ for the clumps with a sharply-peaked NFW profile ($\gamma = 1$) and exponential profile ($\gamma = 1.08$). Since $\theta_{68\%}$ is larger than Fermi's PSF over most of the parameter space the gamma ray emission is generally very extended, but the regions excluded by our point-source and extended-emission analyses nearly overlap, since they both, roughly correspond to brighter gamma-ray emission overall (see \cref{fig:gamma_spectra_constraints_exp}).

Below the lines $d = r_s, \rtidal$ the clump is roughly point-like, and the luminosity increases monotonically with distance as a consequence of CREs losing energy as they propagate. Above this line, and particularly for smaller inner slope, the luminosity depends primarily on scale radius, since CREs on average are produced farther from Earth in the outskirts of the clump.

Assuming an NFW profile, the local density must be at least a factor of $1.4$--$4$  larger than the standard value $\rho_\oplus = \SI{0.3}{\giga\electronvolt/\centi\meter^3}
$ depending on the inner slope, and might be as much $1000$--$3000$ times larger. The corresponding statements are weaker for the exponential profile due to its exponential cutoff, with the minimum density enhancement factor ranging from $\lesssim 0.001$ to $300$--$3000$. Somewhat counter-intuitively, the minimum local density enhancement is largest for the steepest inner slope. This is because gamma-ray flux from a $1^\circ$ cone around the clump is proportional to $d^{-2(1 + \gamma)}$, while the local density enhancement depends more weakly on $\gamma$ as $d^{-\gamma}$. Thus, regions of parameter space with reduced local density enhancement are ruled out by more stringent gamma-ray constraints. \revised{If the line width constraint is relaxed, as discussed in \cref{sec:linewidth}, the effect on \cref{fig:contours_exp_0-74} is to move the gray region slightly to the right. Such a modification does not change these conclusions.}

Surprisingly, such enormous enhancements to the local dark matter density are largely unconstrained. The most stringent limits that apply to the \revised{smallest} scales of interest come from precision measurements of planetary orbits. \citeil{2013AstL...39..141P} find using planetary ephemerides that $\rho<\SI{7.9e8}{\giga\electronvolt/\centi\meter^3}$ at Earth's orbit, and further that the average dark matter density should be no greater than \SI{5.2e3}{\giga\electronvolt/\centi\meter^3} within the orbit of Saturn. This average is still larger than our fiducial $\rho_\oplus$ by a factor of $\sim10^4$, so constraints on the local dark matter density are compatible even with our largest enhancements. Recent constraints based on pulsar timing \citep{Caballero2018} are comparable in strength to those of \citeil{2013AstL...39..141P}.

\revised{On scales of order \SI{1}{\kilo\parsec}, the local dark matter density is generally probed by populations of tracer stars (see e.g. \citeil{Bienayme:2014kva}). Such analyses rely on a model for the Galactic dark matter halo and assume it to be axisymmetric, symmetric above and below the Galactic plane and in dynamical equilibrium. For the nearby clumps we consider these assumptions do not generally hold, and relaxing them can substantially impact the outcomes of these analyses (see e.g. Ref.~\cite{Banik:2016yqm}). Thus, existing stellar constraints cannot be directly applied to the substructures we consider.

In general, o}ur results confirm that a clump must be very nearby in order to source the DAMPE excess\revised{, so any constraints on the local dark matter density must be sensitive to small clumps in order to be relevant for our purposes. However, g}iven what is known about the abundance of dark matter substructures, we can \revised{still} ask whether it is statistically reasonable that such a clump should be found so close by. This is the meaning of the probabilities shown in the bottom-right panels of~\cref{fig:contours_nfw_0-5,fig:contours_nfw_1,fig:contours_exp_0-52,fig:contours_exp_0-74,fig:contours_exp_1-08}.

\begin{figure}
    \centering
    \includegraphics[width=\textwidth]{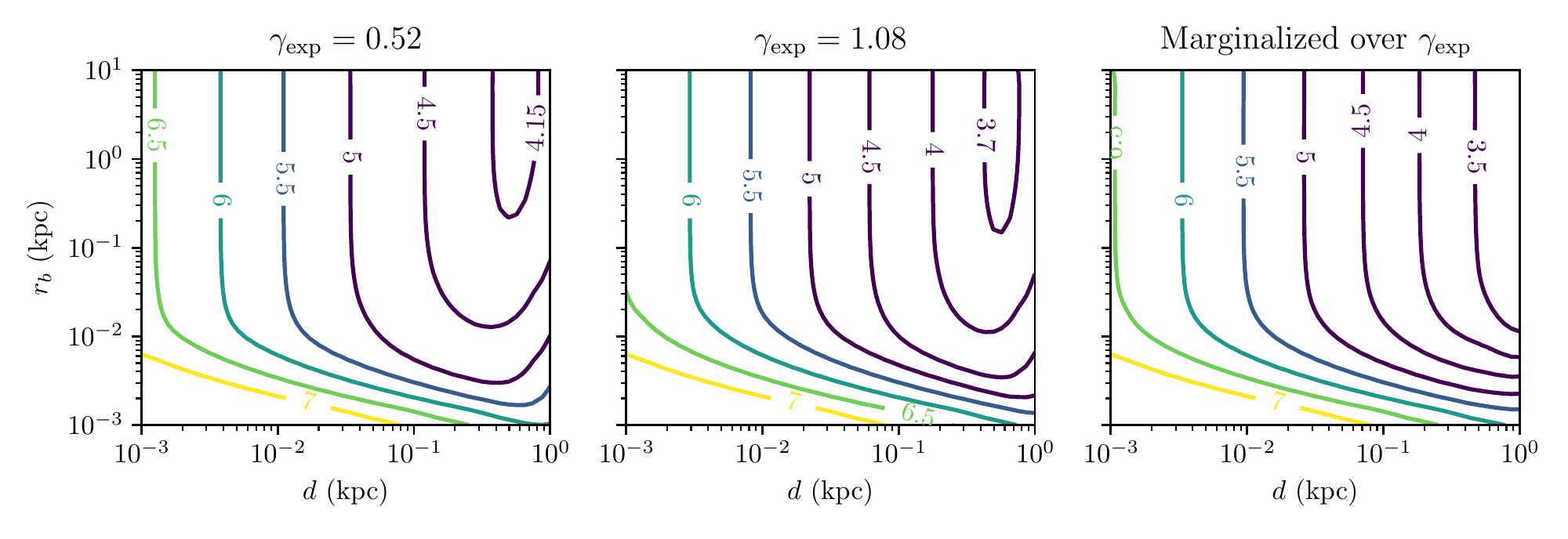}
    \caption{Probability ($z$-score) of having a clump with an exponential profile satisfying the criteria of~\cref{sec:likelihood-criteria}. The left two panels show are for the 25th and 75th percentile values of $\gamma_\u{exp}$ while the right plot marginalizes over $\gamma$ per the distribution of~\cref{eq:gamma-distribution}.}
    \label{fig:exp-likelihoods-all-gamma}
\end{figure}
For the case of an exponential profile, note that we fix $\gamma$ to a set of particular values in our analysis, rather than treating it as a continuously distributed parameter. While this is illustrative, it can make the probabilities difficult to interpret, since a small tail of large $\gamma$ values can contribute significantly to the probability of having a luminous source nearby. Thus, in~\cref{fig:exp-likelihoods-all-gamma}, \revised{we compare probabilities at fixed $\gamma$ with those obtained by marginalizing over $\gamma$ using the distribution of~\cref{eq:gamma-distribution}, as described at the end of \cref{sec:number-densities}. The resulting probabilities are quite similar.}

In each case, the contours show the same basic behavior: at small distances, the probability is nearly linear in the volume, so it increases as $\sim d^3$. At large distances, a clump must be larger or more compact in order to supply the minimum luminosity needed to account for the excess, and since these objects are rarer, the probability begins to drop. Similarly, the probability drops sharply at small radii, since such halos, although more abundant, are rarely sufficiently luminous.

The probabilities themselves are small everywhere, with $z>5$ in much of the parameter space. The fact that the probabilities exceed such a threshold should be regarded with caution: the statistics of small-scale structure is still not fully understood, and further, we have assumed that clumps are uniformly distributed throughout the solar neighborhood. While this is likely an acceptable approximation for our purposes, it is possible that clustering could enhance the probability of being near a clump. It is also possible that the spatial distribution of clumps could be \emph{unfavorable} to the production of a DAMPE-like excess. For example, destruction of clumps by the galactic disk could suppress their local abundance, further reducing the probability of having a very nearby clump.

Again, these probabilities are calculated to give the overall likelihood of producing an observable and line-like indirect detection signal at least as extreme as that observed by DAMPE. To that end, we make cuts that allow for smaller clumps which contribute less to the local overdensity of dark matter. Thus, as discussed in~\cref{sec:likelihood-criteria}, these probabilities cannot be used to estimate the likelihood of a given enhancement to the direct detection rate. Our only aim in this work is to establish the parameter ranges of interest for clumps sourcing line-like indirect detection features. A detailed study of the impact on direct detection must take other factors into account (e.g. the relative velocity of the clump), and we defer this to a later work~\citep{WorkInPrep}.

\begin{figure}
    \centering
    \includegraphics[width=\textwidth]{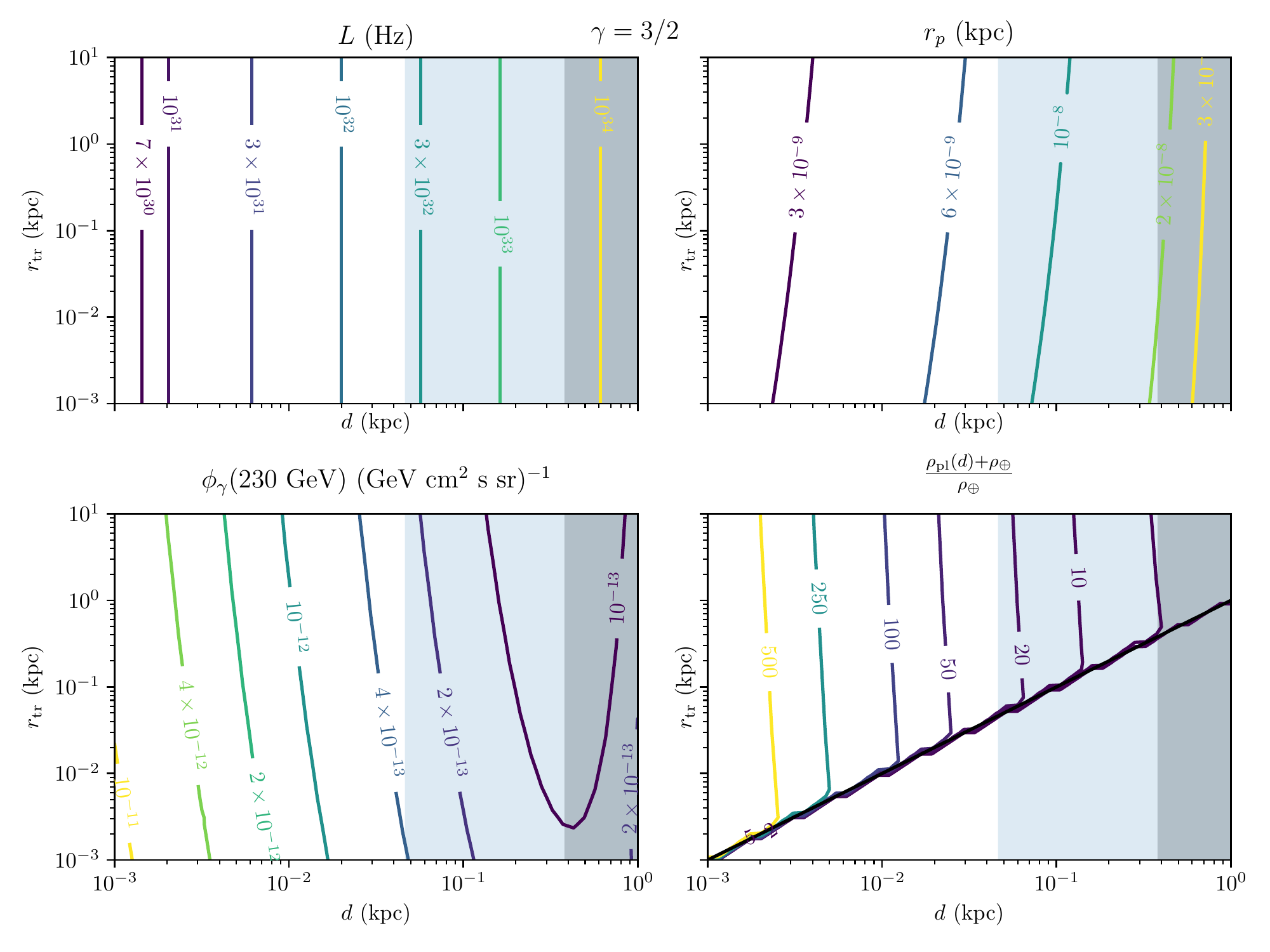}
    \caption{Luminosity, plateau radius, gamma ray flux and relative contribution to the local DM density for a power law clump with index $\gamma = 3/2$. The $y$-axis indicates the clump's truncation radius. The grey region is ruled out since the resulting $e^\pm$ line would be too wide. All points in this plane are excluded by the Fermi gamma-ray point source constraints.}
    \label{fig:contours_pl_3_2}
\end{figure}
\begin{figure}[p]
    \centering
    \includegraphics[width=\textwidth]{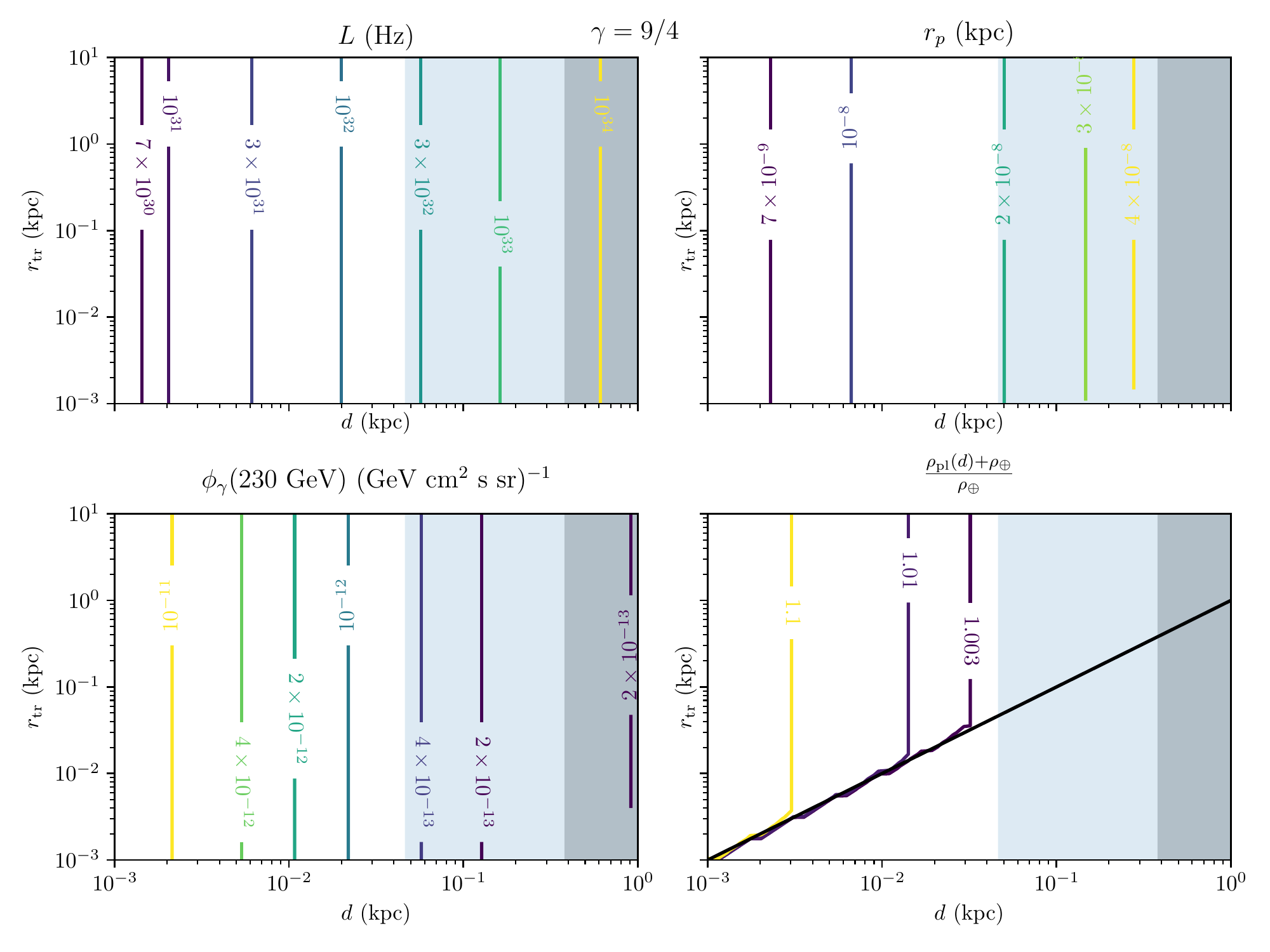}
    \caption{As in \cref{fig:contours_pl_3_2}, but for a power law clump with index $\gamma = 9/4$. Again, as for the $\gamma=3/2$ case shown in \cref{fig:contours_pl_3_2}, all points in this plane are excluded by the Fermi gamma-ray point source constraints.}
    \label{fig:contours_pl_9_4}
\end{figure}

Another possibility is that a nearby clump is not a substructure typical of CDM, but is instead an ultracompact minihalo (UCMH). We consider UCMHs fit to the DAMPE excess with $\gamma = 3/2$ in~\cref{fig:contours_pl_3_2}, which display contours for four UCMH properties in the distance and truncation radius plane:
\begin{itemize}
    \item Luminosity $L$, obtained by assuming the clump is a point-like CRE source;
    \item The gamma-ray flux at \SI{230}{\giga\electronvolt};
    \item The radius of the annihilation plateau (\cref{eq:ucmhprofile}), fixed by setting the clump's luminosity for fixed $(d, r_\u{tr})$ to the values in the first panel;
    \item The local density enhancement from the clump.
\end{itemize}
In the blue region the UCMH would produce an observable CRE anisotropy, and as before the grey region shows the line width constraint. The most important feature of UCMHs is that their gamma-ray emission is point-like, and exceeds the Fermi point-source sensitivity by orders of magnitude, so \emph{the DAMPE excess cannot be produced by UCMHs}. Even if it could, however, the UCMH would have to be within $\sim \SI{4e-2}{\kilo\parsec}$ of Earth.  For $r_\u{tr} < d$, this would generically contribute significantly to the local density, enhancing it by factors greater than 30. We have also considered the steeper index $\gamma = 9/4$ in~\cref{fig:contours_pl_9_4}, and we reach the same conclusion. In this case, the local density enhancement might be smaller than 1\%.

\section{Conclusions}
\label{sec:conclusions}
High-energy cosmic-ray electrons and positrons (CRE) lose energy very efficiently as they propagate in the galaxy via synchrotron and inverse Compton emission. As a result, a narrow, line-like feature in the high-energy portion of the CRE spectrum, such as that tentatively observed by DAMPE, must originate from a nearby source. While astrophysical sources producing nearly monochromatic high-energy CRE might exist, such as cold pulsar wind nebulae, none is close enough to account for a high-energy, bright feature in the CRE spectrum. As a result, if such a feature were ever conclusively detected, a new physics interpretation would appear quite natural and compelling. Among such interpretations, dark matter (DM) annihilation in a nearby, dense clump is perhaps the most plausible.

The existence of a nearby, ``luminous'' (in the sense of \cref{eq:luminosity}) DM clump as a source of a CRE line-like feature such as what reported by DAMPE is constrained by a variety of indirect observations: first, the clump would shine in gamma rays produced by bremsstrahlung off of the high-energy CRE; second, the arrival direction at Earth of CRE would not be isotropic because of the clump proximity, and constraints exist on the level of CRE anisotropy at a given energy; third, given an assumed clump distance and size, one can estimate the probability for the existence of such clump from results of $N$-body simulations.

In this study, we assessed the implications of a CRE line-like feature originating from DM annihilation to electrons and positrons, assuming a set of several functional forms for the clump's assumed spherical density distribution. We considered a set of modified Navarro-Frenk-White density profiles, as well as tidally-truncated exponential profiles motivated directly by numerical simulations and a set of ultracompact profiles with power law indexes of 3/2 and 9/4.

We tested each clump density distribution on the plane defined by the clump distance and ``size'' (defined differently depending on the density profile). On such plane, the distance is bounded from above by the requirement that the feature be sufficiently ``line-like,'' and the size, generally, is bounded from below by gamma-ray constraints. (Depending on the profile, such constraints also bound the distance from below.) Broadly, the most likely clumps from the standpoint of $N$-body simulations have large sizes and distances, but are ruled out by the line width constraint. The region of the plane with highest probabilities and weak observational constraints resides at distances around $0.3$--$\SI{0.4}{\kilo\parsec}$, with clump radii of the same order. We find that such a clump would appear as a very extended gamma-ray source, several 10s of degrees wide in the sky---hundreds of times larger than the Fermi PSF.

The central result of our study is that the most likely clump \revised{parameters compatible with} a DAMPE-like CRE feature would imply a very large enhancement to the local dark matter density\revised{. In \cref{fig:contours_nfw_0-5,fig:contours_nfw_1,fig:contours_exp_0-52,fig:contours_exp_0-74,fig:contours_exp_1-08}, the bottom-right panel shows the likelihood of finding a clump capable of producing the DAMPE excess at a given distance and size, and the middle-right panel shows the local dark matter overdensity in the same plane. The predicted likelihood prefers a region of parameter space with overdensities of order 10. Smaller enhancements on the order of 0.1-1\% are also possible, due to clumps which are both distant and small in size, but these are substantially less likely, being confined to a small sliver of the parameter space. Furthermore, overdensities as large as a factor of 1,000 are compatible with the observational constraints we consider in this work, albeit in tension with the statistical predictions of $N$-body simulations.}
While very large, \revised{even this overdensity} is not constrained by current upper bounds to the local DM density in the Solar System.

We also entertained the possibility that the CRE feature originates from an ultracompact minihalo, with a steep density profile. We find that this possibility is ruled out by gamma-ray constraints. Neglecting those, we predict a local DM density enhancement in this case as well, depending on whether the solar system lies within the clump's truncation radius.

If a CRE line-like feature originates from DM, we are thus led to the conclusion that a significant enhancement to the local DM density is both possible and probable. Direct detection rates would then be enhanced by the same factor as the local density. Additionally, particles in the clump would have a very narrow velocity dispersion, implying that we would live in a stream of cold particles with some overall drift velocity. This would have dramatic implications for the differential event rate at direct detection experiments.

We will explore in a forthcoming study \cite{WorkInPrep} the implications of ``living inside a clump.''

\section*{Acknowledgments}
BVL and SP are partly supported by the U.S.\ Department of Energy grant number DE-SC0010107. We gratefully acknowledge helpful conversations with Piero Madau.

\bibliography{bib}

\end{document}